\documentclass{pas}

\usepackage{array}
\usepackage{multirow}
\usepackage{enumitem}
\usepackage{aas-macros}
\usepackage{orcidlink}
\usepackage{bm}

\makeatletter
\renewcommand{\today}{\two@digits{\the\day}~\two@digits{\the\month}~\the\year}
\makeatother

\begin{document}

\lefttitle{The scale of homogeneity in DESI DR1}
\righttitle{C. Howlett et al.}

\jnlPage{1}{18}
\jnlDoiYr{2026}
\doival{10.1017/pasa.xxxx.xx}

\articletitt{Research Paper}

\title{The scale of homogeneity in DESI Data Release 1}

\author{
Cullan~Howlett\orcidlink{0000-0002-1081-9410}$^{1}$,
L.~Giani$^{2}$,
Elizabeth G.~O'Brien$^{1}$,
A.~Whitford\orcidlink{0000-0001-5829-8637}$^{1}$,
Tamara~M.~Davis\orcidlink{0000-0002-4213-8783}$^{1}$,
J.~Aguilar$^{3}$,
S.~Ahlen\orcidlink{0000-0001-6098-7247}$^{4}$,
U.~Andrade\orcidlink{0000-0002-4118-8236}$^{5,6}$,
E.~Armengaud\orcidlink{0000-0001-7600-5148}$^{7}$,
S.~BenZvi\orcidlink{0000-0001-5537-4710}$^{8}$,
F.~Beutler\orcidlink{0000-0003-0467-5438}$^{9}$,
D.~Bianchi\orcidlink{0000-0001-9712-0006}$^{10,11}$,
D.~Brooks$^{12}$,
R.~Canning$^{13}$,
A.~Carnero Rosell\orcidlink{0000-0003-3044-5150}$^{14,15}$,
F.~J.~Castander\orcidlink{0000-0001-7316-4573}$^{16,17}$,
E.~Chaussidon\orcidlink{0000-0001-8996-4874}$^{3}$,
T.~Claybaugh$^{3}$,
A.~Cuceu\orcidlink{0000-0002-2169-0595}$^{3}$,
A.~de la Macorra\orcidlink{0000-0002-1769-1640}$^{18}$,
Arjun~Dey\orcidlink{0000-0002-4928-4003}$^{19}$,
Biprateep~Dey\orcidlink{0000-0002-5665-7912}$^{20,21}$,
P.~Doel$^{12}$,
D.~Forero-Sánchez\orcidlink{0000-0001-5957-332X}$^{22}$,
J.~E.~Forero-Romero\orcidlink{0000-0002-2890-3725}$^{23,24}$,
E.~Gaztañaga\orcidlink{0000-0001-9632-0815}$^{16,13,17}$,
Satya~{Gontcho A Gontcho}\orcidlink{0000-0003-3142-233X}$^{25}$,
G.~Gutierrez$^{26}$,
C.~Hahn\orcidlink{0000-0003-1197-0902}$^{27}$,
K.~Honscheid\orcidlink{0000-0002-6550-2023}$^{28,29,30}$,
D.~Huterer\orcidlink{0000-0001-6558-0112}$^{31,6}$,
M.~Ishak\orcidlink{0000-0002-6024-466X}$^{32}$,
S.~Juneau\orcidlink{0000-0002-0000-2394}$^{19}$,
T.~Karim\orcidlink{0000-0002-5652-8870}$^{20}$,
D.~Kirkby\orcidlink{0000-0002-8828-5463}$^{33}$,
A.~Kremin\orcidlink{0000-0001-6356-7424}$^{3}$,
O.~Lahav\orcidlink{0000-0002-1134-9035}$^{12}$,
A.~Lambert$^{3}$,
C.~Lamman\orcidlink{0000-0002-6731-9329}$^{30}$,
M.~Landriau\orcidlink{0000-0003-1838-8528}$^{3}$,
L.~Le~Guillou\orcidlink{0000-0001-7178-8868}$^{34}$,
M.~Manera\orcidlink{0000-0003-4962-8934}$^{35,36}$,
A.~Meisner\orcidlink{0000-0002-1125-7384}$^{19}$,
R.~Miquel$^{37,36}$,
J.~Moustakas\orcidlink{0000-0002-2733-4559}$^{38}$,
S.~Nadathur\orcidlink{0000-0001-9070-3102}$^{13}$,
J.~ A.~Newman\orcidlink{0000-0001-8684-2222}$^{21}$,
H.~E.~Noriega\orcidlink{0000-0002-3397-3998}$^{31,5}$,
E.~Paillas\orcidlink{0000-0002-4637-2868}$^{39}$,
N.~Palanque-Delabrouille\orcidlink{0000-0003-3188-784X}$^{7,3}$,
W.~J.~Percival\orcidlink{0000-0002-0644-5727}$^{40,41,42}$,
C.~Poppett$^{3,43,44}$,
F.~Prada\orcidlink{0000-0001-7145-8674}$^{45}$,
I.~P\'erez-R\`afols\orcidlink{0000-0001-6979-0125}$^{46}$,
C.~Ravoux\orcidlink{0000-0002-3500-6635}$^{47}$,
G.~Rossi$^{48}$,
R.~Ruggeri\orcidlink{0000-0002-0394-0896}$^{49}$,
L.~Samushia\orcidlink{0000-0002-1609-5687}$^{50,51}$,
E.~Sanchez\orcidlink{0000-0002-9646-8198}$^{52}$,
C.~Saulder\orcidlink{0000-0002-0408-5633}$^{53}$,
E.~F.~Schlafly\orcidlink{0000-0002-3569-7421}$^{54}$,
D.~Schlegel$^{3}$,
M.~Schubnell$^{31,6}$,
J.~Silber\orcidlink{0000-0002-3461-0320}$^{3}$,
M.~Siudek\orcidlink{0000-0002-2949-2155}$^{17,15}$,
G.~Tarl\'{e}\orcidlink{0000-0003-1704-0781}$^{6}$,
B.~A.~Weaver$^{19}$,
and H.~Zou\orcidlink{0000-0002-6684-3997}$^{55}$
\\
}

\affil{Affiliations are listed at the end of the manuscript}

\corresp{C. Howlett; email: c.howlett@uq.edu.au}

\citeauth{C. Howlett et al, The scale of homogeneity in DESI Data Release 1, {\it Publications of the Astronomical Society of Australia} {\bf 00}, 1--18. https://doi.org/10.1017/pasa.xxxx.xx}

\history{(Received \today; revised xx xx xxxx; accepted xx xx xxxx)}

\begin{abstract}
We present measurements of the scale of cosmic homogeneity using the first data release (DR1) of the Dark Energy Spectroscopic Instrument (DESI), the largest publicly available spectroscopic extragalactic survey to date. Using a counts-in-spheres estimator applied to nearly 5 million galaxies and quasars across six independent redshift bins spanning $0.1 < z < 2.1$, we measure the correlation dimension $D_2(r)$ and define the homogeneity scale $R_H$ as the point at which the galaxy distribution reaches within 1\% of the homogeneous asymptotic value. We find that all tracers clearly transition to homogeneity, with the galaxy homogeneity scale ranging from $93.8^{+0.9}_{-0.7}$ to $142.6^{+1.4}_{-1.4}$ $h^{-1}$Mpc depending on the tracer redshift, galaxy bias, and modelling used to extract the scale. Using a cosmology-dependent model, we reconstruct the underlying matter homogeneity scale, finding it decreases monotonically with redshift in excellent agreement with $\Lambda$CDM predictions. We identify a systematic offset of up to $\sim 10\%$ between measurements in the North and South Galactic Caps, largely due to fiber assignment effects reproduced in our simulations. Accounting for this, and with a complementary demonstration using toy models that our methodology correctly identifies inhomogeneous universes in all but rather finely-tuned observer configurations, this hence provides strong evidence in favour of the cosmological principle across our full redshift range. Our matter homogeneity scale measurements are consistent with, and substantially more precise than, previous constraints from WiggleZ and SDSS, demonstrating the power of DESI data for precision tests of the cosmological principle.
\end{abstract}


\maketitle

\section{INTRODUCTION}
\label{sec:introduction}

Despite its simplicity, the standard cosmological model describes remarkably well a broad range of independent cosmological observations at various scales. This success relies on a set of assumptions about the Universe's energy content, its geometrical structure, and the nature of gravity. Specifically, it postulates the existence of an indirectly observed dark sector consisting of Cold Dark Matter (CDM) and dark energy in the form of a Cosmological Constant ($\Lambda$); and of a spacetime background satisfying the cosmological principle and evolving according to the equations of General Relativity. 

Thanks to the astonishing accuracy reached by modern cosmological surveys, these assumptions and the standard model can nowadays be tested with unprecedented precision. For example, recent evidence has emerged challenging the interpretation of dark energy as a cosmological constant, hinting that some sort of dynamical dark energy is preferred by the data \citep{DES2024,DESIDR2II,DESI2024VII,DESI2024VI}, though this preference remains weak enough that we cannot rule out the standard cosmological model. In this work we focus on testing the validity of a different facet of the standard model, the cosmological principle. This posits that the Universe, on sufficiently large scales, is statistically homogeneous and isotropic, implying that its statistical properties are rotationally and translationally invariant.

Since the average density in the Universe is of order $\sim 10^{-30} \rm{g}/\rm{cm}^3$, Cartesian arguments\footnote{`Je pense, donc je suis', Descartes, \textit{Discourse on the method}, 1637.} of course suggest that these assumptions must break down at smaller scales, giving rise to the diverse structures we see around us. To quantify and measure at which scale the cosmological principle breaks down is hence an important task of modern cosmology, both for its theoretical and observational implications. On the theoretical side, it is still unclear whether the complex web of cosmic structures emergent in the late Universe can alter its evolution; scenarios which fall under the umbrella-term of `cosmological backreaction'. Some studies suggest that this, rather than `Dark Energy', could be responsible for the observed accelerated expansion of the Universe (see e.g. \citealt{Rasanen2004,Alnes2006,Biswas2007,Buchert2008,Wiltshire2009,Racz2017,Heinesen2020a}), whereas others claim that it should be completely negligible \citep{Ishibashi2006,Green2014,Green2016,Macpherson2019}. In between these opposite scenarios, other analyses have found that backreaction effects can have an appreciable impact on our cosmological inference and might be causing some observational inconsistencies in constraints on the $\Lambda$CDM model (see e.g. \citealt{Buchert2003,Ben-Dayan2012,Ben-Dayan2014,Fleury2013,Bagheri2014, Rubart2014,Buchert2015,Bolejko2016,Wijenayake2016,Anton2023,Clifton2024,Giani2024,Giani2025}). From the observational point of view, it is important to quantify at which scales our cosmological inference can be biased by environmental effects.

For these reasons, several methods and techniques have been developed to measure the transition to homogeneity, as reflected by the large literature on the subject. Constraints on the anisotropy of the early Universe from CMB, X-ray and radio observations, combined with the Copernican principle (the assumption that we do not live in a special place in the Universe), can be used to infer the homogeneity scale (e.g. \citealt{Scharf2000,Blake2002a,Camacho-Quevedo2022,Shao2025b}). Perhaps the most powerful probe of cosmic homogeneity in the late Universe is the 3-dimensional distribution of galaxies. The number of neighbouring galaxies in a sphere of radius $r$, if their distribution is homogeneous, should be proportional to the sphere volume. The scale at which this proportionality is reached, $R_H$, can be used to characterize the transition to homogeneity  \citep{Martinez1994,Martinez1998,Amendola1999,Pan2000,Yadav2005,Sarkar2009,SylosLabini2010,SylosLabini2011,Scrimgeour2012,Ntelis2017,SylosLabini2025}. Several estimators have been considered in the literature to measure this scale while mitigating possible biases \citep{Borgani1995,Martinez2002,Shao2025a,Giani2026}, and several studies have proposed using the homogeneity scale to test cosmological models \citep{Ntelis2018,Avila2022,Shao2024}.

In this paper we produce new measurements of this homogeneity scale, defined as the scale at which deviations from homogeneity, as estimated from the correlation dimension, are below the percent level. We do this by applying a count-in-spheres based estimator to the first Data Release (DR1) of the Dark Energy Spectroscopic Instrument (DESI) collaboration. The DESI DR1 data, with more than 30 million measurements of galaxy and quasar spectra, is the largest publicly-available spectroscopic extragalactic sample to date, offering a unique opportunity to measure the homogeneity scale and its redshift evolution with unprecedented precision. Similar measurements using this data were recently presented in \cite{SylosLabini2025}, although, as we will discuss in this work, our approach is very different, as are our conclusions.

The layout of this paper is as follows: in Section~\ref{sec:desi} we review the DESI instrument, the data product and the tools for its validation. In Section~\ref{sec:corr} we review the correlation dimension as an estimator of cosmic homogeneity, its theoretical foundation and its robustness, measuring it in toy models, simulations and the real data. In Section~\ref{sec:rh} we present our measurement of the homogeneity scale using both cosmology-dependent and independent models of the correlation dimension and compare it with previous results. Finally, Section~\ref{sec:conclusion} is devoted to a discussion of our results. Throughout this work, where needed, we assume a fiducial cosmology equal to the primary cosmology used in the \texttt{AbacusSummit} suite of simulations \citep{Maksimova2021}, with cosmological parameters $\omega_{\mathrm{b}}=0.02237$, $\omega_{\mathrm{cdm}}=0.1200$, $h=0.6736$, $A_{\mathrm{s}}=2.0830\times10^{-9}$, $n_{\mathrm{s}}=0.9649$, and two massless + one massive neutrino with energy density $\omega_{\nu}=0.00064420$. 

\section{THE DARK ENERGY SPECTROSCOPIC INSTRUMENT}
\label{sec:desi}

The Dark Energy Spectroscopic Instrument (DESI; \citealt{DESI2022}) is a robotic, fiber-fed, optical spectrograph mounted on the Mayall 4m telescope at Kitt Peak National Observatory, capable of simultaneously observing $5,000$ objects \citep{DESI2016II,Miller2024,Poppett2024}. Since May 2021, the instrument has been carrying out an ambitious 8-year program to obtain spectra for 63 million galaxy and quasars over $17,000\,\mathrm{deg}^{2}$ of sky \citep{Guy2023,Schlafly2023}, with the first year of data having been publicly released (DR1; \citealt{DESIDR1}).

The main goals of the survey include new precision measurements of the clustering of the Universe's large-scale structure as traced by galaxies, quasars and the Lyman-$\alpha$ forest. Such measurements using the DR1 data \citep{DESI2024II} were subsequently used to produce novel constraints on the time-varying nature of dark energy using Baryon Acoustic Oscillations (BAO) and Redshift-Space Distortions (RSD; \citealt{DESI2024VII, DESI2024VI, DESI2024IV, DESI2024III, DESI2024V}), in combination with data from the Cosmic Microwave Background \citep{Planck2020VI} and Type Ia supernovae \citep{Brout2022, DES2024, Rubin2025}. These provide tentative hints that the equation of state of dark energy is not constant with time, a result which has remained persistent with the analysis of DESI's 3-year dataset (DR2; \citealt{DESIDR2II,DESIDR2I}), but remains below the level at which we can rule out $\Lambda$CDM.

\subsection{Data Release 1}
\label{sec:dr1}
Beyond DESI's bread-and-butter science goals, this wealth of data offers a unique opportunity to investigate fundamental assumptions in our analysis of cosmological data, such as the cosmological principle. For this purpose, we make use of DESI's DR1 dataset, which allows us to probe the transition to homogeneity, and extract a homogeneity scale for the Universe, across a wide range of epochs with a single instrument. Due to the nature of target selection \citep{Myers2023}, DESI's primary targets are divided into four main tracers consisting of the Bright Galaxy Sample (BGS), Luminous Red Galaxies (LRGs), Emission Line Galaxies (ELGs) and Quasars (QSOs). For each of these, large-scale structure catalogues were produced \citep{Ross2025} that provide robust, cosmology-ready sets of coordinates and redshifts with weights to account for systematics arising from imaging variations, fiber collisions, and redshift failures. Random, unclustered catalogues matching the survey footprint, redshift distribution and data selection functions are also produced as part of that work, as well as so-called `FKP' weights \citep{Feldman1994} that minimise the variance in two-point clustering statistics measured from data with redshift-dependent number density.

\begin{figure*}[]
\centering
    \includegraphics[scale=0.6]{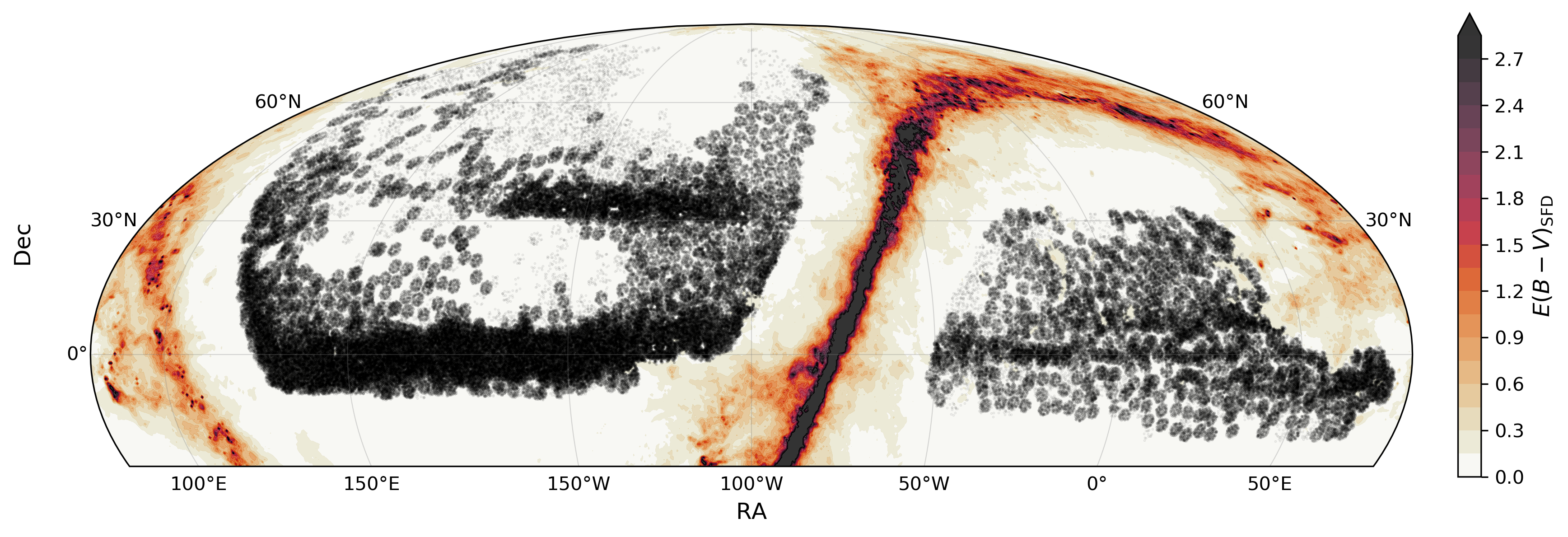}
	\caption{The sky distribution of the DESI DR1 data used in this work (black points), compared to the distribution of dust in the Milky Way (coloured by E(B-V) as measured by \citealt{Schlegel1998}). The data at positive/negative Galactic latitudes (i.e., on the left/right-hand sides of the plot) are referred to as the North and South Galactic Caps (NGC/SGC) respectively. The patchiness of the DR1 data, and individual DESI tiles, are readily apparent, requiring us to make use of the random catalogues and systematics weights provided with the DESI large-scale structure catalogues when estimating the homogeneity scale.} 
 \label{fig:desi_sky}
\end{figure*}

\begin{figure}[]
\centering
    \includegraphics[scale=0.7]{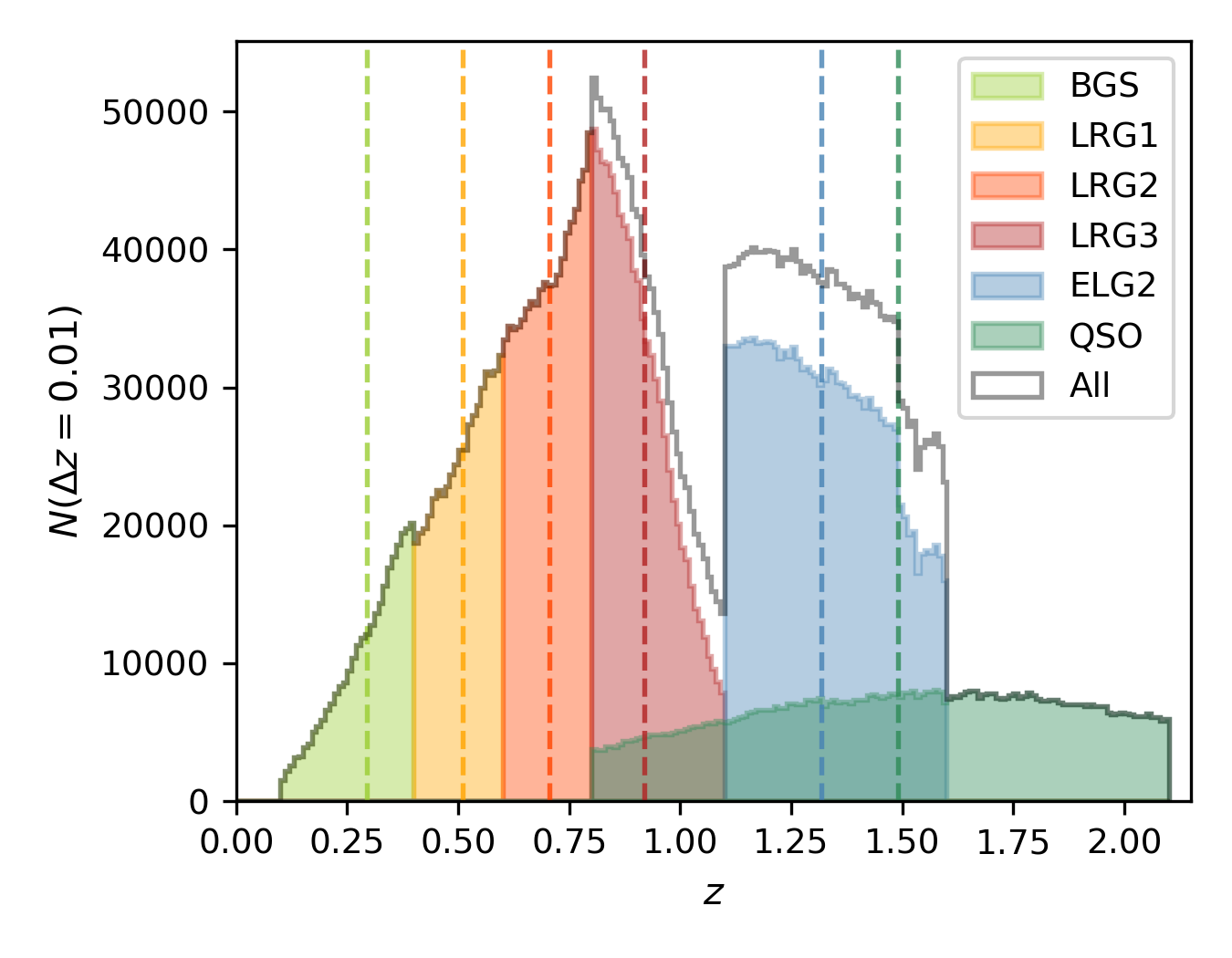}
	\caption{The number of galaxies as a function of redshift in the DESI DR1 data used in this work, binned in bins of width $\Delta z = 0.01$. Different colours represent different samples (which will be used consistently throughout this work), with characteristics detailed further in Table~\ref{tab:data}. The vertical dashed lines indicate the effective redshift of each sample, which will be used as the redshift at which the homogeneity scale is reported.} 
 \label{fig:desi_nz}
\end{figure}

We make use of all of these existing data products in this work, splitting the four tracers into the same redshift bins as used for the DR1 BAO and RSD analyses, spanning $0.1<z<2.1$ with an effective skyarea of between $\sim5,700-7,500\,\mathrm{deg}^{2}$ depending on the tracer. The data is split into North and South Galactic Cap regions (NGC and SGC respectively), denoting galaxies above and below the plane of the Milky Way. Figure~\ref{fig:desi_sky} shows the sky coverage of the data, while Figure~\ref{fig:desi_nz} shows the number of galaxies in our sample as a function of redshift, highlighting the contributions from each tracer and the `effective' redshift of each of our bins.

Due to the overlap between the LRG and ELG tracers between $0.8<z<1.1$, the weak expected constraining power of the ELGs in this redshift range in DR1, and the additional complexity required to combine these two tracers, we only use the LRGs in this redshift range.\footnote{Putting it another way, we only use the `LRG3' bin defined in the original DR1 BAO and RSD analyses, and do not consider the additional/alternative `ELG1' or `LRG3+ELG1' measurement bins.} This results in 6 effectively independent redshift bins with which we can measure the homogeneity scale.\footnote{While the QSO sample also overlaps with both the LRG3 and ELG2 samples, \cite{DESI2024VI} estimate the correlation between the ELG2 and QSO two-point clustering measurements to be less than 10\%, and even smaller between LRG3 and QSO. As such, they treat them as independent for the purposes of constraining cosmological parameters. We apply the same arguments here.} Table~\ref{tab:data} lists the most important characteristics of each of these bins. 

\begin{table}
 \caption{Characteristics of the 6 independent galaxy samples used in this work, constructed from DESI DR1 data. Columns denote how we refer to the sample, the redshift range, effective sky area, number of galaxies, and effective redshift.}\label{tab:data}
  \centering
 {\tablefont\begin{tabular}{@{\extracolsep{\fill}}lcccccc}
   \toprule
    \multirow{2}{*}{Name} & Redshift & \multicolumn{2}{c}{Sky Area ($\mathrm{deg^{2}})$} & 
     \multicolumn{2}{c}{$N_{\mathrm{gals}}$} & \multirow{2}{*}{$z_{\mathrm{eff}}$}\\
      & Range & NGC & SGC & NGC & SGC & \\
     \hline
     
    BGS & $0.1 < z < 0.4$ & 5,300 & 2,173 & 217,614 & 82,429 & 0.295 \\
    LRG1 & $0.4 < z < 0.6$ & 3,756 & 1,984 & 346,761 & 160,510 & 0.510 \\
    LRG2 & $0.6 < z < 0.8$ & 3,756 & 1,984 & 533,955 & 237,939 & 0.706 \\
    LRG3 & $0.8 < z < 1.1$ & 3,756 & 1,984 & 595,419 & 264,403 & 0.919 \\
    ELG2 & $1.1 < z < 1.6$ & 3,888 & 2,036 & 1,054,840 & 360,867 & 1.317 \\
    QSO & $0.8 < z < 2.1$ & 4,644 & 2,605 & 555,913 & 300,918 & 1.491
   \botrule
    \end{tabular}}
\end{table}

\subsection{Mock galaxy catalogues}
\label{sec:mocks}
In order to validate our methodology for testing homogeneity, explore systematics, and evaluate the statistical uncertainties on our measurements, we use two sets of simulations produced alongside the DESI DR1 large-scale structure datasets, which mimic the most important aspects of the data. These are described in detail in \cite{DESI2024III}, so only the most salient points will be summarized here. Both sets of simulations use the same fiducial cosmology as provided at the end of Section~\ref{sec:introduction}.

\subsubsection{Abacus-2}
The first set of simulations are 25 so-called `altmtl Abacus-2' realisations for each tracer. These represent our most accurate representations of the data, replicating the DR1 sky and redshifts distributions, as well as the impacts of fiber-assignment and the subsequent systematic weights applied to the data to partially mitigate these effects. 

The backbone of these are the high-resolution \texttt{AbacusSummit} N-Body simulations \citep{Garrison2018, Maksimova2021}, within which halos have been identified using the \texttt{COMPASO} code \citep{Bose2022, Hadzhiyska2022}, and populated with mock galaxies using Halo Occupation Distribution models tuned to the DESI Early Data Release data \citep{Alam2020, Rocher2023, Smith2024, Yuan2024}. The mocks are then run through the DESI \texttt{fiberassign} code to assign fibers to the simulated galaxies based on the exact configuration files (specifying hardware configuration, sky fibres, standard stars, etc.) used at the time the real data were observed. This hence reproduces the on-sky selection and observing effects in the real data \citep{Lasker2025}. 

Due to their high-fidelity, these are the mocks used to test for systematics in our homogeneity scale measurement and fitting methodology before applying it to the real data. As the mocks are, by definition, $\Lambda$CDM universes which exhibit homogeneity and isotropy on large-scales and contain the same galaxy bias and selection function as the data, we also use these as comparison points for the real data --- for instance, if we see that the data prefers a different homogeneity scale to the `altmtl Abacus-2' simulations, we know that this must be due to differences in underlying cosmology or unknown systematics, as opposed to data incompleteness or the way in which galaxies populate the dark matter field.

\subsubsection{EZmocks}
\label{sec:ezmocks}
Our second set of simulations are 1000 EZmock realisations for each tracer. These are approximate N-Body simulations, using the \texttt{EZmock} code \citep{Chuang2015a}, that can be run significantly faster than full N-body simulations. The dark matter fields are then populated directly using an effective bias model \citep{Cheng2021} before being run through an approximate fast fiber-assignment (FFA) process \citep{Bianchi2025}. The FFA method avoids the computational cost of simulating the full observing loop by emulating the probability of a galaxy being observed based on the number of neighbouring galaxies that would compete for a fiber. 

Compared to the altmtl Abacus-2 mocks, these EZmocks exhibit weaker fiber-collision effects (less reduction in clustering amplitude on small scales), and an overall lower clustering amplitude (which may be linked to either the fiber-collision effects or EZmock N-Body method itself). As such, we use these only for estimating the statistical errors on our measurements, including a rescaling factor in the measured covariance to account for the lower clustering amplitude. These rescaling factors, applied directly to the covariance matrix, are given in Table~\ref{tab:rescaling} and match those used in the BAO and RSD analyses of \cite{DESI2024III} and \cite{DESI2024V}.

When fitting any measurements in this work, we apply \cite{Hartlap2007} and \cite{Percival2014} correction factors to the EZmock covariance, based on the use of 1000 simulations and the number of free parameters in the fitting model, to account for bias and error underestimation arising from using a finite number of simulations to estimate the true covariance matrix.

\begin{table}
 \caption{Multiplicative factors applied to the EZmock-based covariance matrix estimates for each DESI DR1 sample to account for small differences in the clustering amplitude and fibre-collisions between the EZmocks and data. These values match those used in the BAO and RSD analyses of \cite{DESI2024III} and \cite{DESI2024V}.}
 \label{tab:rescaling}
 \centering
 {\tablefont\begin{tabular}{@{\extracolsep{\fill}}cccccc}
   \toprule
     BGS & LRG1 & LRG2 & LRG3 & ELG2 & QSO \\
     \hline
    1.39 & 1.15 & 1.15 & 1.22 & 1.29 & 1.11
   \botrule
    \end{tabular}}
\end{table}

\section{THE CORRELATION DIMENSION}
\label{sec:corr}
A common method for investigating the transition to homogeneity and homogeneity scale in the Universe is using `scaled counts-in-spheres' (e.g., \citealt{Hogg2005, Sarkar2009, Scrimgeour2012, Ntelis2017}). In essence, one draws spheres of increasing radius around each tracer in the sample, and counts the number of neighboring tracers in each sphere, scaled by a random catalogue that cancels out the effect of survey boundaries and selection effects. The average of these counts as a function of the radius defines the \textit{mean scaled counts} $N(r)$. 

In a homogeneous distribution, the mean counts-in-spheres scales with the volume, $\propto r^3$. The mean \textit{scaled} counts, being the ratio of a data catalogue and random catalogue, will hence tend to a constant if the data is homogeneous. The radius $r$ at which $N(r)$ reaches some desired threshold value can be used to define a homogeneity scale. However, therein lies a problem --- there is a degeneracy between the asymptotic value of the mean scaled counts and the unknown mean density of the data used to estimate it (otherwise known as the integral constraint), which results in ambiguity in choosing a threshold value that can be translated between datasets \citep{Scrimgeour2012,Heinesen2020b}. For this reason, a more reliable estimator of the homogeneity scale is the correlation dimension $D_2$ \citep{Borgani1995}
\begin{equation}
    D_2(r) \equiv 3 + \frac{d \ln N(r)}{d \ln r}\;,
    \label{eq:D2}
\end{equation}
which measures only the scaling of $N(r)$, and therefore is insensitive to the mean value of the density of tracers. $D_{2}$ measured from a homogeneous distribution will tend to 3. As with the mean scaled counts, the scale of homogeneity can be defined as the point at which $D_{2}(r)$ crosses some threshold value, typically set to be 2.97 (i.e., within 1\% of the homogeneous asymptotic value).\footnote{Note however that there do exist specific inhomogeneous distributions where the correlation dimension asymptotes to 3 while the mean scaled counts diverge, an example being $N(r) \propto r^3\ln(1+r)$. As such, while $D_{2}$ is preferred for determining the \textit{scale} of homogeneity, both $N(r)$ and $D_{2}(r) $ should be inspected to test for homogeneity in general.}

The mean scaled counts $N(r)$ (and hence $D_{2}(r)$) are also straightforwardly related to the monopole of the correlation function $\xi_{0}(s)$ \citep{Peebles1980} 
\begin{equation}
    N(r) = 1+\frac{3}{r^{3}}\int_{0}^{r}s^{2}\xi_{0}(s)\,ds = 1+\bar{\xi}(r),
    \label{eq:msc}
\end{equation}
where we have defined the volume-averaged correlation function $\bar{\xi}(r)\equiv\frac{3}{r^{3}}\int_{0}^{r}s^{2}\xi_{0}(s)\,ds$. Substituting this into Equation~\ref{eq:D2}
\begin{equation}
    D_{2}(r) = 3\frac{1+\xi_{0}(r)}{1+\bar{\xi}(r)},
    \label{eq:D2_2}
\end{equation}
which allows this and $N(r)$ to be modeled using the same techniques as for the correlation function or power spectrum.

By virtue of this relationship, the correlation dimension and homogeneity scale clearly carry information on the clustering of galaxies and matter. For example, \cite{Ntelis2018} and \cite{Shao2024} have investigated its potential as a standard ruler, while \cite{Avila2022} have explored how measurements of the homogeneity scale at different redshifts can probe the linear growth rate $f(z)$. This latter case can be derived by defining the scale of homogeneity in the matter field, $R^{\mathrm{matter}}_{H}$, as the solution to the equation $D_{2}(R^{\mathrm{matter}}_{H}) = 3(1-\epsilon)$, where $\epsilon$ is a constant (typically taken in the literature as 0.01 as discussed previously). Substituting into Equation~\ref{eq:D2_2} and assuming linear theory for the matter correlation function allows us to write
\begin{equation}
    1+D^{2}(a)\xi(R_{H},a=1) = (1-\epsilon)(1+D^{2}(a)\bar{\xi}(R_{H},a=1))\approx1-\epsilon,
\end{equation}
where the last approximation makes use of the fact that, at the homogeneity scale, $N(r) = 1+D^{2}(a)\bar{\xi}(R_{H},a=1)\approx 1$.

If we now take the time derivative of this expression, we find that $\frac{\partial}{\partial a}\left[D^{2}(a)\xi(R_{H},a=1)\right]=0$, which can be solved to find
\begin{equation}
    f(a) = \frac{d \ln D(a)}{d \ln a} = - \frac{1}{2}\frac{d\ln R_{H}}{d \ln a}\frac{d\ln \xi(r, a=1)}{d \ln r}\biggl|_{R_{H}},
    \label{eq:growth}
\end{equation}
i.e., the growth rate of structure can be related to the time derivative of the homogeneity scale normalised by the gradient of the $z=0$ matter correlation function.

\subsection{Estimation from data}
\label{sec:estimator}
In order to estimate the mean scaled counts and correlation dimension from data, we use a modified version of the estimator from \cite{Scrimgeour2012}
\begin{equation}
    \hat{N}(r) = \alpha \sum_{i=1}^{G}w^{g}_{i}\left[\frac{\sum_{j=1}^{G}w^{g}_{j}\Theta(r - r_{ij} )}{\sum_{j=1}^{R}w^{r}_{j}\Theta(r - r_{ij})}\right],
    \label{eq:estimator}
\end{equation}
where $G$ and $R$ are the total number of galaxies and random points in our dataset, and 
\begin{equation}
    \alpha = \frac{\sum_{j=1}^{R}w^{r}_{j}}{\left[\sum_{j=1}^{G}w^{g}_{j}\right]^{2}}
\end{equation}
normalises the counts while accounting for the relative numbers of random points to galaxies. $\Theta(r-r_{ij})$ is the Heaviside function, equal to one if the separation between galaxy $i$ and the $j^{\mathrm{th}}$ galaxy or random (denoted by $r_{ij}$) is less than the sphere radius $r$, and zero otherwise. Finally, $w^{g}$ and $w^{r}$ are the weights assigned to the galaxies and random points, which are the product of the systematics weight for that data point, and the FKP weight. Overall, this expression is directly comparable to Equation 4 in \cite{Scrimgeour2012} except for the generalization to include weights.

In order to estimate the mean scaled counts, we first need to convert the Right Ascension, Declination and redshift of each galaxy/random to cartesian coordinates, which we do using our fiducial cosmology. We then make use of a fast K-d tree algorithm to identify all neighbours for galaxy $i$ within our maximum sphere radius and generate a cumulative histogram in narrow radius bins to evaluate the inner weighted sums in Equation~\ref{eq:estimator}. This is iterated over all galaxies to perform the outer weighted sum. The correlation dimension is then produced by fitting a cubic spline to $\ln(N(r))$ as a function of $\ln(r)$ and evaluating its derivative at the midpoint of each of our radial bins. Statistical uncertainties on our measurements are obtained by applying the same procedure to the 1000 EZmock realisations (for each tracer) introduced in Section~\ref{sec:ezmocks} and computing the covariance matrix.

\subsection{Toy models}

The estimator for the mean scaled counts in Section~\ref{sec:estimator} requires two notable steps: 1) a conversion from angular coordinates and redshifts to cartesian coordinates using a fiducial cosmology; and 2) the use of a random catalogue to normalise the counts. This may raise the question of to what extent the estimator and associated homogeneity scales can actually be used to detect inhomogeneous or anisotropic distributions without circularity. 

To test this, we apply our estimator to four different toy models, each of which is generated by populating a box of edge-length 2$h^{-1}$Gpc with $10^{6}$ particles. These are:
\begin{enumerate}[label=\alph*)]
\item{a homogeneous and isotropic uniform distribution;}
\item{an isotropic (about coordinate [0,0,0]) but inhomogeneous Gaussian distribution with standard deviation 400$h^{-1}$Mpc;}
\item{an isotropic (about coordinate [0,0,0]) but inhomogeneous Giant Void distribution evacuated using a Generalized Gaussian probability $P(r)\propto e^{-(r/\sigma)^4}$ with scale $\sigma=800h^{-1}$Mpc;}
\item{an inhomogeneous and anisotropic `Swiss Cheese' fractal model produced by randomly populating a uniform distribution with completely empty spherical voids whose radii are drawn from a probability distribution $P(r)\propto r^{-3}$. This distribution is fractal as a result of its statistical self-similarity and scale-independence, looking identical regardless of how one `zooms in' on the structure.}

\end{enumerate}
In analysing each of these cases, we compute the mean scaled counts centred on two different observers (at [0,0,0] $h^{-1}\mathrm{Mpc}$ and [-200,-200,-200] $h^{-1}\mathrm{Mpc}$ respectively) with two different maximum redshift cuts to ensure our results are not complicated by the edges of the simulation box. We also use two types of random catalogue placed in the same box with tens times the number of points (prior to our redshift cuts being applied), setting the weights $w_{g}=w_{r}=1$. Our first random catalogue is uniform across the volume of the box. Our second is initially uniform, but the redshifts of the random points are `reshuffled' by picking randomly from the data so that the random redshifts capture the data's radial selection function. This is commonly done for large-scale structure analyses, such as in the Sloan Digital Sky Survey Data Release 12 \citep{Reid2016} and DESI \citep{DESI2024II}. As we will demonstrate, this is an important difference, with strong implications for whether or not a given distribution will be identified as homogeneous \citep{SylosLabini1998,SylosLabini2025}. This second approach still keeps the \textit{angular} distributions of the random points uniform --- this is also in line with standard large-scale structure analyses, which typically aim for a uniform selection function across the sky and so only apply small-scale completeness masks to the randoms to account for holes due to intervening stars, `veto' regions of poor quality photometric/spectroscopic data, and account for fiber collisions.

\begin{figure}[]
\centering
    \includegraphics[scale=0.65]{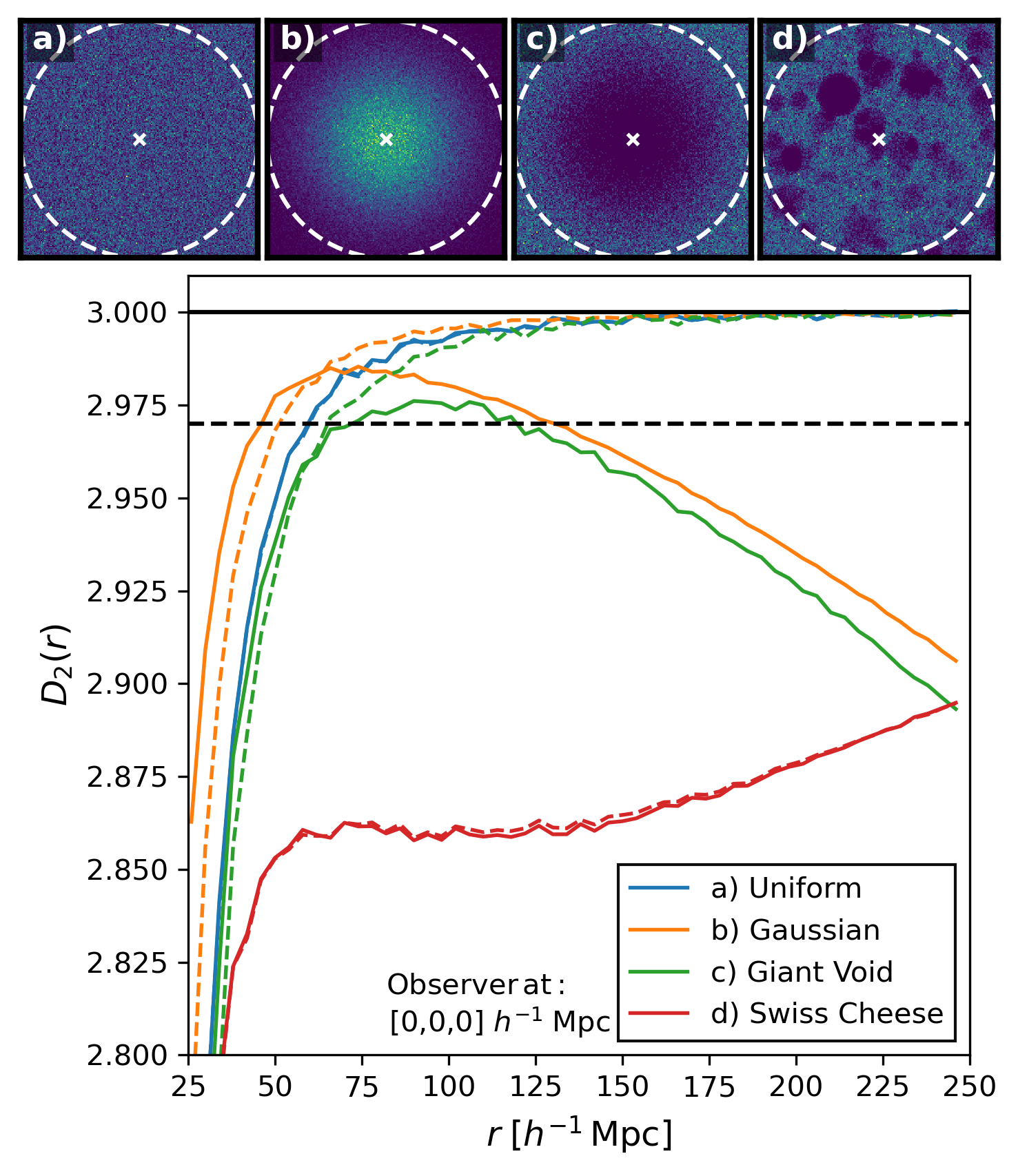}
    \includegraphics[scale=0.65]{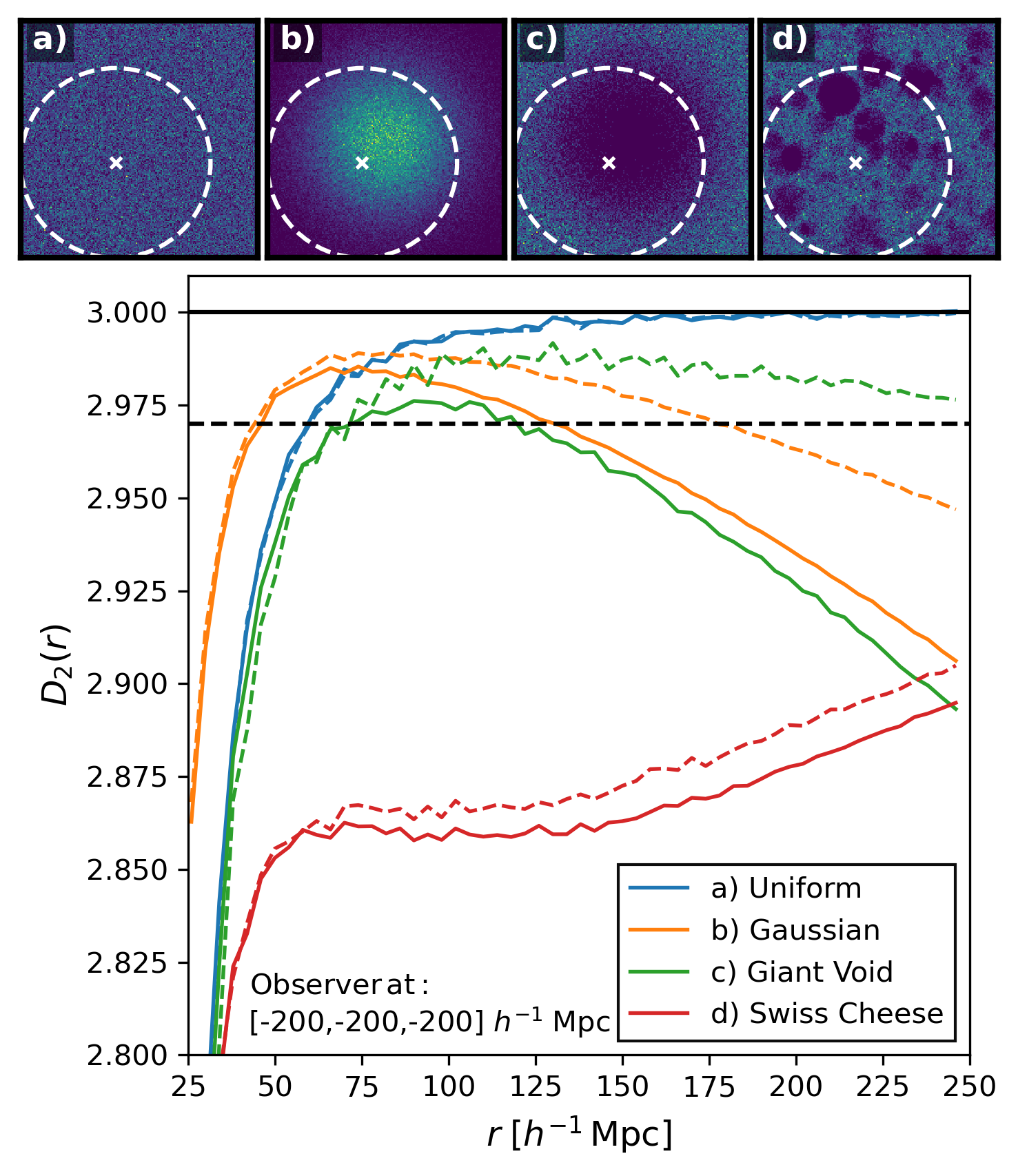}
	\caption{Measurement of the correlation dimension as a function of scale for 4 different toy models, constructed by populating a box of edge-length 2$h^{-1}$Gpc with $10^{6}$ particles and measuring them using the estimator described in Section~\ref{sec:estimator}. The top and bottom figures shows measurements for observers placed at [0,0,0] $h^{-1}$Mpc and [-200,-200,-200] $h^{-1}$Mpc respectively. In both cases, the top inset panels show 200$h^{-1}$Mpc thick slices of the four distributions as a visual aid, where the white cross and circle shows the observer's location and maximum distance cut used for the measurements. The solid lines show measurements using a fully uniform random catalogue. Dashed lines are produced using a `reshuffled' random catalogue where the redshifts of the random points are generated by drawing from the data. The uniform and `Swiss Cheese' models are clearly delineated as homogeneous and inhomogeneous respectively regardless of the choice of observer or random catalogue. The isotropic but inhomogeneous Gaussian and Giant Void models however can appear homogeneous depending on the observer's proximity to the centre of the distribution when a `reshuffled' random catalogue is used (as is done in our real DESI data).}
 \label{fig:toy_models}
\end{figure}

The resulting correlation dimensions are shown in Figure~\ref{fig:toy_models}, alongside visualisations of each of the distributions. Consider first the results when the observer is placed in the center of the simulation (top panel). We can see that when a uniform random catalogue is used, only the homogeneous and isotropic uniform distribution exhibits a correlation dimension that crosses the 1\% homogeneity threshold and asymptotes to the expected value of 3 within the scales measured (which corresponds to $1/8^{\mathrm{th}}$ of the box size). The inhomogeneous Gaussian and Giant Void distributions both cross the 1\% threshold, and approach homogeneity, but then diverge soon after once the sphere radius becomes a significant fraction of the characteristic scale of these two distributions. The inhomogeneous and anisotropic `Swiss Cheese' model remains far below the homogeneity scale on all scales measured. Overall, this demonstrates that in the case where the selection function and redshift evolution of the data can be modelled \textit{a priori}, the correlation dimension and our estimator of it perform well in distinguishing homogeneous and inhomogeneous distributions on samples with similar size and volume to our DESI data.

However, let us now consider the case using the reshuffled randoms. In this scenario, we see that for an observer placed in the box centre the isotropic Gaussian and Giant Void simulations mistakenly appear homogeneous (note the dashed lines in the top panel of Figure~\ref{fig:toy_models}). This is a pernicious problem, as raised in \cite{SylosLabini2025} --- in their work they instead consider the ideal case of a volume limited sample, such that one could be sure that the data selection is purely uniform (and if there is negligible source evolution across the sample). However, we argue that for any realistic sample it is effectively impossible to distinguish between a radial selection function in the data and a true intrinsic radial distribution in the Universe. This hence makes it impossible to demonstrate whether the chosen sample is truly volume limited. The result is that the choice to analyze with \textit{or} without normalisation via a reshuffled random catalogue will affect the results, with neither case obviously superior --- in the case with normalisation, homogeneity may be recovered even for inhomogeneous distributions; in the case without normalisation, homogeneous distributions are likely to look inhomogeneous because what appears to be a volume-limited sample is actually subject to some radial selection. In this scenario, the correlation dimension measured using counts-in-spheres centered on galaxies is simply an insufficient statistic. 

There are several cases, however, that we argue justify using our estimator of the correlation dimension and reshuffled randoms rather than attempting to restrict to volume limited samples. First, in Figure~\ref{fig:toy_models} it is interesting to see that the results for the fractal `Swiss Cheese' universe are not affected by the use of a random catalogue -- even when the random redshift distribution is forced to match the data, the correlation dimension is unchanged. Secondly, from the bottom panel of Figure~\ref{fig:toy_models} we see that for an observer that does not lie in the centre of the Gaussian or Giant Void distributions our estimator of the correlation dimension is still able to identify these as inhomogeneous --- the curve does not asymptote to $3$ regardless of whether uniform or reshuffled randoms are used. This in essence arises because the data now exhibits significant anisotropy in its angular distribution which is not present in the reshuffled random catalogue. Interestingly, the results for the `Swiss Cheese' model remain effectively independent of the observer's location.

From these results, we conclude that our use of the correlation dimension normalised by a reshuffled random catalogue is preferred over trying to construct a volume limited sample from the current DESI data. This is based on the following arguments: 1) complexities in survey selection functions (for DESI or otherwise) make it quite likely that an apparent volume-limited sample is actually not so; 2) it is quite \textit{unlikely} that in an intrinsically inhomogeneous Universe, any observer would measure an apparently isotropic distribution of sources, such as those living in the centres of the spherical toy models considered here; and 3) our estimator is still sufficient to rule out or confirm fractal universes even with reshuffled randoms. The results in this work should hence be considered as a direct test for evidence of fractal or anisotropic universes, but we cannot rule out isotropic but inhomogeneous universes. However, it is important to note that homogeneity measurements using 3D galaxy clustering, as in this work, have an advantage over similar measurements from the CMB or using X-ray clusters in that they are sensitive to anisotropy at any redshift within the sample, rather than only at a particular redshift, or integrated across the full redshift range \citep{Laurent2016}.

\subsection{DESI data}
\label{sec:desi_msc}
\begin{figure}[]
\centering
    \includegraphics[scale=0.8]{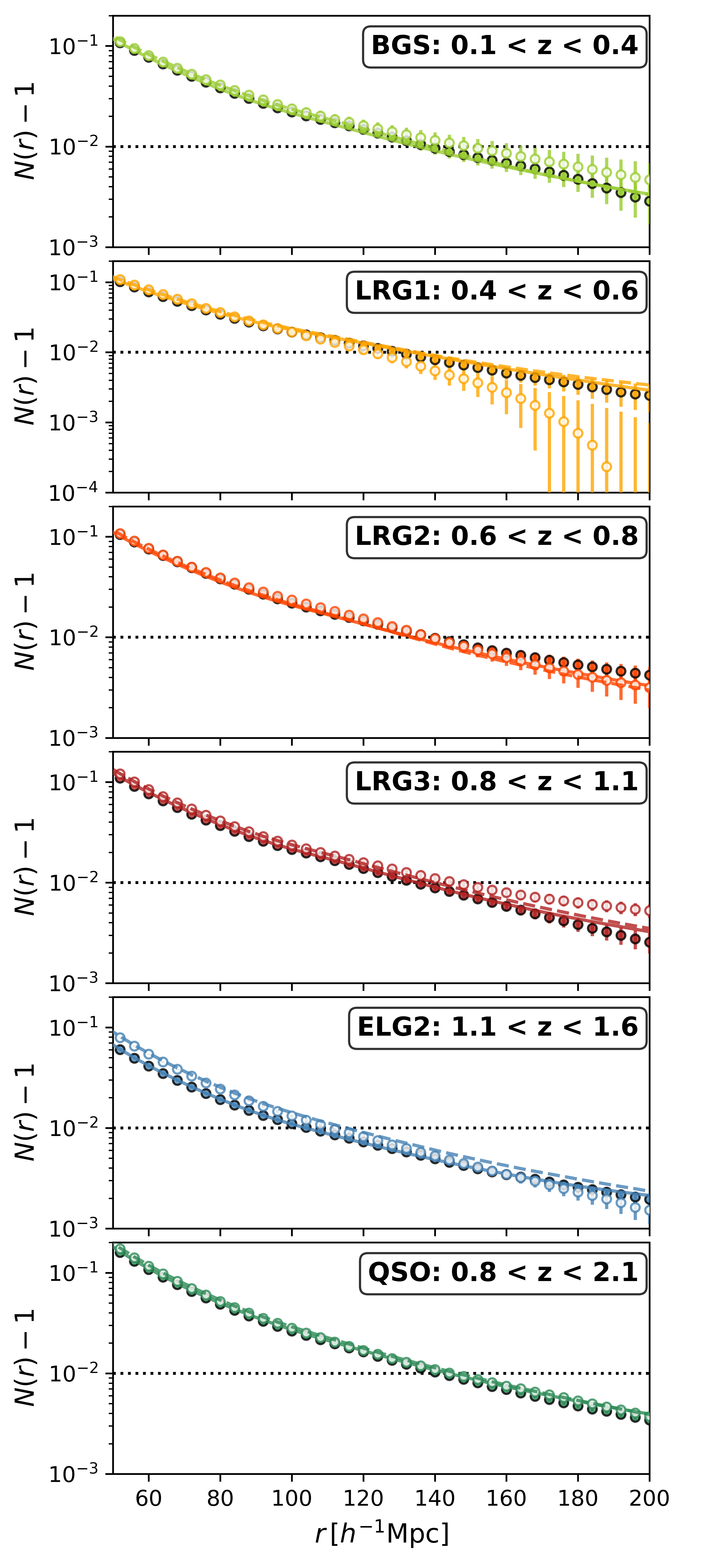}
	\caption{Measurements of the mean scaled counts as a function of scale for each of the 6 DESI redshift bins considered in this work and both NGC and SGC sky areas (closed and open points respectively). Also shown are similar measurements for the Abacus-2 mocks, averaged over all 25 realisations, for both the NGC (solid lines) and SGC (dashed lines). The dotted horizontal line demarcates a $1\%$ deviation from the homogeneous expectation to guide the eye. In all cases we see that the data converges towards homogeneity on scales $\gtrsim 100h^{-1}$Mpc in a similar fashion as the $\Lambda$CDM-based mocks, but there are notable differences between the various tracers, sky areas, and between the mocks and data, which are investigated further using the correlation dimension.} 
 \label{fig:msc_data}
\end{figure}

We next apply our estimator to the real DESI data and mocks detailed in Sections~\ref{sec:dr1} and~\ref{sec:mocks}. We measure all 6 tracer redshift bins and both the NGC and SGC separately to investigate the consistency in the measurements between these hemispheres. Figure~\ref{fig:msc_data} shows the mean scaled counts for the real data in each of these cases. For all tracers and both NGC and SGC we see that the mean scaled counts converge towards 1, indicating a convergence to homogeneity on scales $\gtrsim 100h^{-1}$Mpc. We do see some differences between the rate of convergence between the tracers, with the ELG2 sample converging at the smallest scales, and the QSO sample at the largest. The tracers also show noticeable differences between the NGC and SGC, beyond just a simple amplitude offset, which we demonstrate translates to differences in the correlation dimension between NGC and SGC. On top of the data we show the mean of the Abacus-2 mocks --- these exhibit similar convergent behaviour, but better agreement between NGC and SGC, and differ significantly from the data on large scales for many of the tracers.

\begin{figure*}[]
\centering
    \includegraphics[scale=1.0]{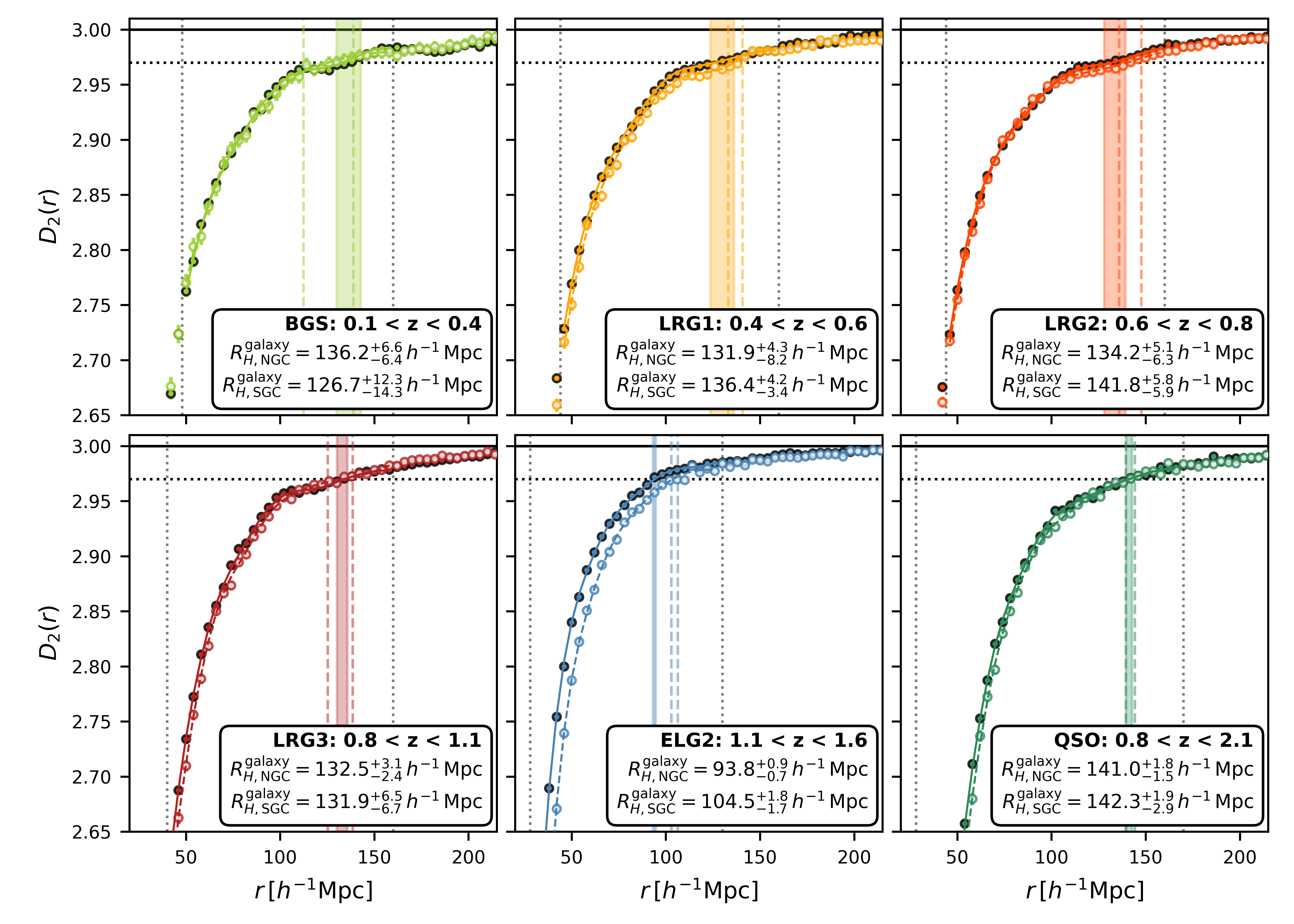}
	\caption{Measurements of the correlation dimension as a function of scale for each of the 6 DESI redshift bins considered in this work for the NGC (closed points) and SGC (open points). Alongside the measurements we show the best-fit basis-spline models (solid lines for NGC and dashed lines for SGC), used to extract the quoted homogeneity scale $R^{\mathrm{galaxy}}_{H}$ for that galaxy sample. The constraints on this scale are highlighted as the colored vertical band and pair of dashed lines for the NGC and SGC respectively, and also included as text. The black dotted vertical lines show the range of scales included in our fits, discussed in Section~\ref{sec:rh}, while the dotted and solid horizontal lines denote the values of $D_{2}(r)=2.97$ (1\% homogeneity threshold) and the asymptotic homogeneity value $D_{2}(r)=3.0$ respectively. The homogeneity scale is defined as the point where the data crosses the dotted horizontal line. We see that all tracers and sky patches approach the expectations for homogeneous Universe, although with differences in the homogeneity scale between tracers and between sky patches for the same tracer.} 
 \label{fig:cd_data}
\end{figure*}

To investigate this further, we turn to the correlation dimension. These measurements are shown in Figure~\ref{fig:cd_data} for the 6 tracer bins and both sky caps. In each case, we still see a clear transition to homogeneity, reaching within the 1\% threshold on scales $\sim90 - 140h^{-1}$Mpc depending on the exact sample (with the ELG2 bin being the most homogeneous and the QSO sample the least). However, as was suspected based on the mean scaled counts, there are significant differences in the shapes of the correlation dimension between the NGC and SGC except for the BGS sample. Another interesting feature, most prominent in the LRG samples, is that of the BAO bump at around 100$h^{-1}$Mpc --- the presence of this feature so close to the threshold for homogeneity demonstrates that this is typically a weak deviation from homogeneity, but one that is nonetheless detectable at high significance in DESI.

As foreshadowed from the additional information presented in Figure~\ref{fig:cd_data}, the exact shapes of the correlation dimension for each tracer and sky cap lead to differences in the inferred homogeneity scale. We discuss how these measurements are obtained and their interpretation in the next section.

\section{THE HOMOGENEITY SCALE}
\label{sec:rh}
Given our measurements of the correlation dimension in the DESI data, we seek to evaluate a `homogeneity scale' --- a point at which the data crosses the threshold value (1\%) of the asymptotic value for homogeneity that can subsequently be used to compare between different redshifts and sky patches, and potentially as a cosmological probe.

To do this, we take two approaches. The first is a basis-spline fit to the data, which can subsequently be used to provide a measure of the crossing-point satisfying $D_{2}(R_{H})=2.97$ without assuming a cosmological model for the shape of the correlation dimension. Our second, model-dependent, approach uses a full-shape model for the monopole of the galaxy power spectrum which is then transformed into the correlation dimension using Equation~\ref{eq:D2_2} via the correlation function. This second approach requires both an assumed cosmological model to set the shape of the matter correlation dimension, plus a bias model to relate the matter and galaxy distributions. Although model-dependent, the benefits of this second approach are that the bias parameters can be fit to the data and then used to rescale the correlation dimension back to what would be expected for the matter field at that redshift and for that cosmological model, effectively removing the complex effects of galaxy bias and providing a measurement that can be more easily translated between datasets.

 The range of fitting scales used for each tracer differ slightly, balancing the need for our cosmology-dependent model to include large enough scales in the fit to capture the homogeneity scale for that set of \textit{galaxies}, but also small enough scales that the model is a good fit to the data around the corresponding \textit{matter} homogeneity scale for extrapolation --- all while ensuring the model itself has enough flexibility to capture the small scale shape of the correlation dimension. This was particularly tricky for the highly biased, high redshift QSO sample, where the homogeneity scale in the galaxies is particularly large, but where the matter homogeneity scale was the lowest of our samples. The appropriate range of scales was hence arrived at iteratively, and was then also used for our model-independent fits. The values are given in Tables~\ref{tab:model-independent} and~\ref{tab:model-dependent}, and shown as dotted lines in Figs.~\ref{fig:cd_data} and \ref{fig:matter_homogeneity_NGC}.  Overall, we found that our results are robust to different choices of radial binning and fitting ranges as demonstrated in Appendix A.

\subsection{Model-independent results}
We start by discussing our basis-spline fits, which are independent of an assumed, cosmology dependent, shape for the correlation function or correlation dimension.\footnote{Note however that there is still an assumed fiducial cosmology used to convert our galaxy positions and redshifts into cartesian coordinates, which affects cosmological interpretation of the homogeneity scale (e.g., \cite{Ntelis2018}. An alternative would be to use instead the angular scale of homogeneity as in e.g., \cite{Andrade2022}.}   The best-fit curves for each of the data samples were previously shown in Figure~\ref{fig:cd_data}, alongside the constraints on the homogeneity scale. The polynomial order for the basis-spline was chosen to be cubic (degree 3) and the number of spline knots/coefficients was chosen by dividing the number of data points within our fitting range by 3 (rounded down). We found that instead choosing spline knots per every 2 or 4 data points gave less than $1\sigma$ variation in the measured homogeneity scale. The homogeneity scale was identified by solving $D_{2}(R_{H})=2.97$ using numerical root finding and taking the solution with the smallest $r$. Our data is generally constraining enough that each posterior sample has only a single crossing point. However, in the presence of noisy data and a BAO feature close to the homogeneity scale, we find that picking the first crossing point in the presence of multiple crossings leads to an enlarged uncertainty on the homogeneity scale, where some of the posterior samples have a crossing point before the BAO scale and others after. This reflects a true ambiguity in the homogeneity scale due to the BAO feature that is correctly captured in our uncertainties.

Results for both data and Abacus-2 mocks were obtained using the \texttt{emcee} MCMC package \citep{ForemanMackey2013}, and are summarised in Table~\ref{tab:model-independent}.

\begin{table*}
 \caption{Constraints on the galaxy homogeneity scale from model-independent, basis-spline fits to the correlation dimension. The first three columns show the name, effective redshift, and fitting range for each tracer sample. We then show homogeneity scale constraints for the NGC and SGC sky patches for the Abacus-2 mock mean (with covariance matrix divided by 25 to reflect the error on the mean) and real data. For the data, we also provide the best-fit $\chi^{2}$ value and degrees of freedom, demonstrating good fits for all our tracers. The last two columns show the absolute difference in the homogeneity scale between the data and homogeneous $\Lambda$CDM Abacus-2 mocks, normalised by the uncertainty on each measurement. We see good agreement between the constraints from the data and mocks, indicating that for all tracers and sky patches the data reaches homogeneity at a similar point to the homogeneous $\Lambda$CDM expectations, and that differences between the NGC and SGC results are consistent with fiber assignment and survey selection effects.}\label{tab:model-independent}
  \centering
 {\tablefont\begin{tabular}{@{\extracolsep{\fill}}lcccccccccc}
   \toprule
    \multirow{3}{*}{Name} & \multirow{3}{*}{$z_{\mathrm{eff}}$} & \multirow{3}{*}{Fitting Range} & \multicolumn{2}{c}{Abacus-2} & \multicolumn{4}{c}{Data} & \multicolumn{2}{c}{\multirow{2}{*}{$|\Delta R_{H}^{\mathrm{galaxy}}|$}}\\
      & & & \multicolumn{2}{c}{$R_{H}^{\mathrm{galaxy}}\,(h^{-1}\mathrm{Mpc})$} & \multicolumn{2}{c}{$R_{H}^{\mathrm{galaxy}}\,(h^{-1}\mathrm{Mpc})$} & \multicolumn{2}{c}{$\chi^{2}$/dof} & &  \\
      & & & NGC & SGC & NGC & SGC & NGC & SGC & NGC & SGC \\
     \hline
    BGS & 0.295 & $48h^{-1}\mathrm{Mpc} < r < 160h^{-1}\mathrm{Mpc}$ & $130.3^{+1.7}_{-1.2}$ & $137.2^{+2.7}_{-1.3}$ & $136.2^{+6.6}_{-6.4}$ & $126.7^{+12.3}_{-14.3}$ & $17.0/23$ & $14.4/23$ & $0.89\sigma$ & $0.85\sigma$\\
    LRG1 & 0.510 & $44h^{-1}\mathrm{Mpc} < r < 160h^{-1}\mathrm{Mpc}$ & $130.9^{+1.1}_{-1.2}$ & $132.5^{+1.3}_{-1.2}$ & $131.9^{+4.3}_{-8.2}$ & $136.4^{+4.2}_{-3.4}$ & $9.1/24$ & $29.8/24$ & $0.12\sigma$ & $1.07\sigma$\\
    LRG2 & 0.706 & $44h^{-1}\mathrm{Mpc} < r < 160h^{-1}\mathrm{Mpc}$ & $130.0^{+0.7}_{-0.8}$ & $134.3^{+1.2}_{-1.0}$& $134.2^{+5.1}_{-6.3}$ & $141.8^{+5.8}_{-5.9}$ & $8.6/24$ & $17.7/24$ & $0.66\sigma$ & $1.25\sigma$ \\
    LRG3 & 0.919 & $40h^{-1}\mathrm{Mpc} < r < 160h^{-1}\mathrm{Mpc}$ & $133.0^{+0.6}_{-0.7}$ & $138.6^{+0.8}_{-1.1}$ & $132.5^{+3.1}_{-2.4}$ & $131.9^{+6.5}_{-6.7}$ & $21.1/24$ & $13.8/24$ & $0.16\sigma$ & $1.02\sigma$ \\
    ELG2 & 1.317 & $28h^{-1}\mathrm{Mpc} < r < 132h^{-1}\mathrm{Mpc}$ & $94.9^{+0.2}_{-0.1}$ & $105.1^{+0.5}_{-0.4}$ & $93.8^{+0.9}_{-0.7}$ & $104.5^{+1.8}_{-1.7}$ & $33.6/21$ & $18.7/21$ & $1.21\sigma$ & $0.33\sigma$ \\
    QSO &  1.491 & $28h^{-1}\mathrm{Mpc} < r < 172h^{-1}\mathrm{Mpc}$ & $141.4^{+0.3}_{-0.4}$ & $144.6^{+0.7}_{-0.3}$ & $141.0^{+1.8}_{-1.5}$ & $142.3^{+1.9}_{-2.9}$ & $31.6/28$ & $28.4/28$ & $0.22\sigma$ & $1.20\sigma$
   \botrule
    \end{tabular}}
\end{table*}

The results for the real data confirm the qualitative conclusions made in Section~\ref{sec:desi_msc}. The homogeneity scale as measured using the DESI galaxies in each of our tracer bins varies between $93.8_{-0.7}^{+0.9}h^{-1}$Mpc and $142.3_{-2.9}^{+1.9}h^{-1}$Mpc, with a relative uncertainty of between $\sim1\%$ and $10\%$ depending on the tracer and sky patch. The $\chi^{2}$ per degree-of-freedom for our fits are all reasonable, indicating a robust fit.

The ELG2 sample clearly exhibits the smallest homogeneity scale, and the QSO and LRG2 samples the largest. The ELG2 and QSO samples are the most precisely constrained --- despite their relatively low number density compared to the BGS and LRG samples, these two samples cover the largest cosmological volume, and the uncertainty around the homogeneity scale, being a large scale measurement, is dominated primarily by cosmic variance rather than shot-noise. Hence, the best constraints come from the tracers that cover large cosmic volumes, even if that coverage is somewhat sparse. A more formal derivation of the expected variance in the correlation dimension can be seen in \cite{Yadav2010}, which demonstrates that (under the assumptions of Gaussianity and ignoring survey geometry) this can be related to the usual combination of $P(k)+1/\bar{n}$ that also determines the accuracy of BAO and RSD analyses, but integrated over spatial scales in such a way as to reduce the impact of shot-noise compared to these analyses.

Also included in Table~\ref{tab:model-independent} are fits to the Abacus-2 mock means, using the EZmock covariance divided by 25 (for averaging over 25 simulations) as the standard error on each data point. These provide a $\Lambda$CDM comparison to the data in each redshift bin and galactic cap, and show that the visible NGC/SGC differences in Figs~\ref{fig:msc_data} and~\ref{fig:cd_data} translate into a shift in the fitted homogeneity scale between the two sky patches, in both the mocks and the data.

In the Abacus-2 mocks, the SGC homogeneity scale is consistently larger than NGC, by an amount that varies by tracer, despite the simulations sharing a single $\Lambda$CDM cosmology and consistent galaxy bias prescription. Repeating our methodology on ``complete'' ELG2 mocks -- catalogues built with the real DESI footprint (excluding areas of poor input imaging, high extinction, bright stars, or low-quality tiles) but without fiber assignment or collisions -- shows this is largely a fiber-assignment effect: the weighted mean NGC$-$SGC offset falls from $-9.63\pm0.46\,h^{-1}\mathrm{Mpc}$ in the altmtl mocks to $-0.46\pm0.12\,h^{-1}\mathrm{Mpc}$ once fiber assignment is removed. Because $D_2$ and the mean scaled counts (Eqs.~\ref{eq:D2_2} and~\ref{eq:estimator}) depend on the \textit{integrated} correlation function, the small-scale pair deficit caused by fiber collisions propagates to large scales; the systematics weights, designed to correct the large-scale correlation function, cannot fully compensate for it, particularly given the larger small-scale completeness variations in SGC than NGC \citep{DESI2024II}.

Because this systematic is present in the mocks, we do not combine the NGC and SGC measurements, and instead compare each sky patch to the $\Lambda$CDM predictions individually; the resulting deviations are given in the last two columns of Table~\ref{tab:model-independent}. In every case the homogeneity scale remains consistent with $\Lambda$CDM despite the NGC/SGC offset.

\subsection{Model-dependent results}
We now turn to model-dependent fits. These can be used to reconstruct the corresponding homogeneity scale in the combined dark and baryonic matter field, providing a more standard baseline to compare with results from other datasets and potentially use as a cosmological consistency check (i.e., for the growth rate of structure using Equation~\ref{eq:growth}).

Our model is constructed by evaluating the redshift-space power spectrum multipoles using the \texttt{LPTVelocileptors} module in the \texttt{desilike} code.\footnote{https://github.com/cosmodesi/desilike} \texttt{desilike} is the official codebase used for the key DESI BAO and RSD results, while the \texttt{LPTVelocileptors} model uses a combination of Lagrangian Perturbation and Effective Field Theory approaches to model the redshift-space clustering of biased tracers including additional counter- and stochastic-terms to account for non-linear physics that cannot be modelled perturbatively. \cite{Chen2020a, Chen2020b} describe this model in detail, while \cite{Maus2025a, Maus2025b} demonstrate the excellent accuracy with which this model can recover cosmological parameters from DESI data. The full model is rather lengthy, consisting of a number of terms evaluated using integrals over the linear power spectrum, and so won't be repeated here. However, \cite{Maus2025a} (Eqs. 3.5 and 3.6 therein) provide a neat summary of the power spectrum model, highlighting the dependence on the galaxy bias terms ($b_{1}$, $b_{2}$, $b_{3}$, $b_{s}$), counterterms ($\alpha_{n}$), and stochastic terms ($\mathrm{SN}_{n}$).

Once the power spectrum has been specified, this is converted to a correlation function using a Fourier transform. The monopole is then used to compute the mean scaled counts and correlation dimension via Eqs.~\ref{eq:msc} and~\ref{eq:D2_2}. As for the model-independent fits, the homogeneity scale is then defined as the point satisfying $D_{2}(R_{H})=2.97$. As the model for the correlation dimension depends only on the monopole, this restricts the number of necessary counter and stochastic terms. Furthermore, in our fits we also fix the cosmology of the model to only find the best-fitting bias, counter, and stochastic parameters necessary to reconstruct the matter correlation dimension. Lastly, we follow the practice of \cite{Maus2025a} and fix $b_{3}=0$, finding that the remaining parameters are sufficient to provide a good enough fit to all our tracers on the scales of interest. Overall, we are left with 5 free parameters: the physical linear, quadratic, and stochastic bias parameters plus a single pair of counter and stochastic terms; $[b_{1}\sigma_{8}$, $b_{2}\sigma^{2}_{8}$, $b_{s}\sigma^{2}_{8}$, $\alpha_{0}$, $\mathrm{SN}_{0}]$. These are all given wide, flat priors, and the bias parameters are normalised relative to $\sigma_{8}$, the variance in spheres of radius $8h^{-1}$Mpc, at the effective redshift of each tracer. The correlation dimension for matter, or unbiased tracers, is recovered by setting $b_{1}\sigma_{8}=\sigma_{8}$ and $b_{2}\sigma^{2}_{8}=b_{s}\sigma^{2}_{8}=\alpha_{0}=\mathrm{SN}_{0}=0.0$.

\begin{figure}[]
\centering
    \includegraphics[scale=0.8]{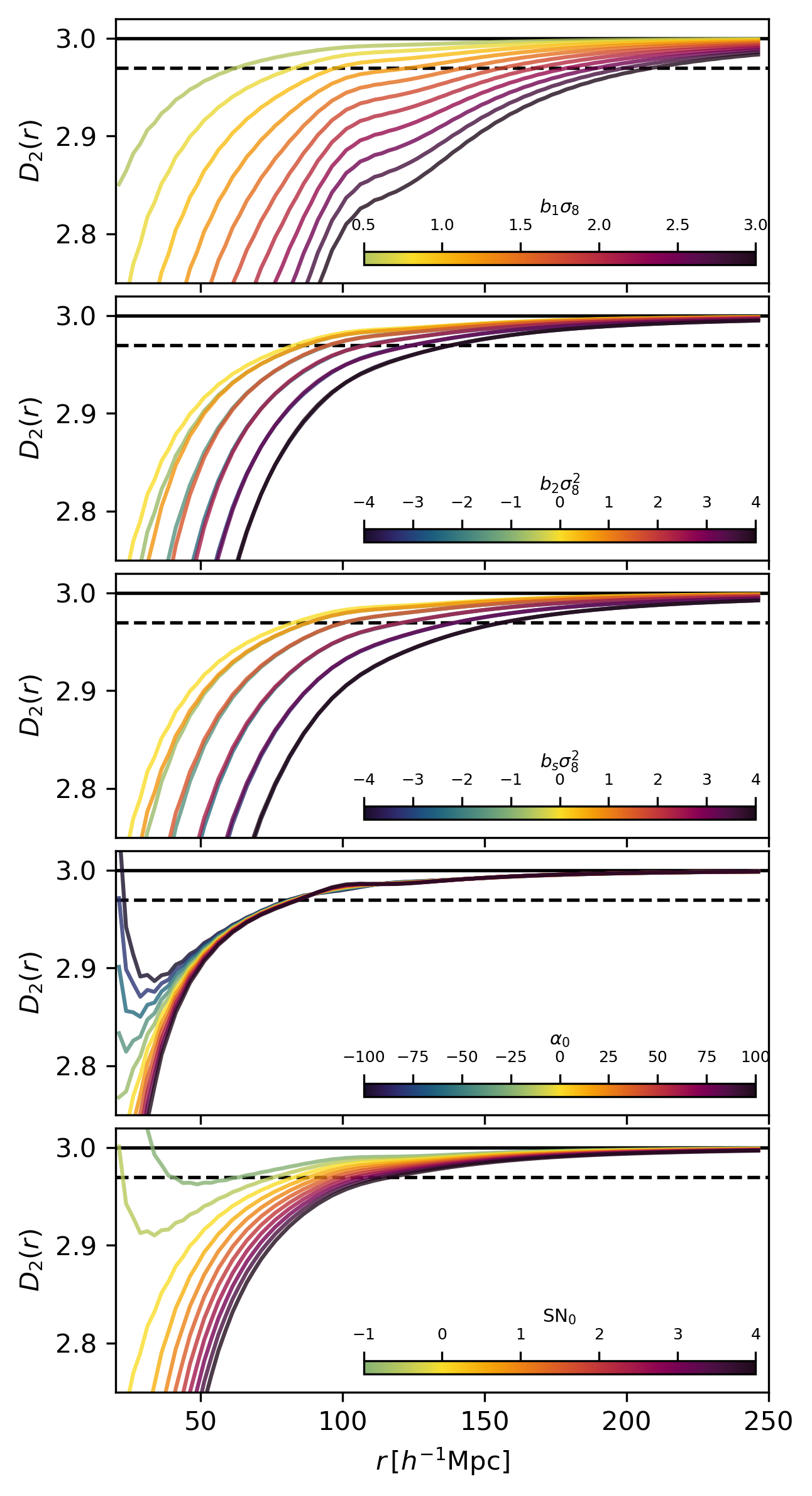}
	\caption{The correlation dimension at $z=0$ for our fiducial cosmology as a function of different free parameters in our cosmology-dependent \texttt{LPTVelocileptors} model. In each panel, we vary only a single free parameter: one of the linear, quadratic, or stochastic galaxy biases ($b_{1}\sigma_{8}$, $b_{2}\sigma^{2}_{8}$, $b_{s}\sigma^{2}_{8}$), the EFT counterterm ($\alpha_{0}$), or the stochastic term ($\mathrm{SN_{0}}$), and keep the remaining parameters fixed at values corresponding to the correlation dimension for matter. In general, larger amplitude galaxy bias values increase the homogeneity scale by shifting the correlation dimension to larger scales, while the counter and stochastic terms change the non-linear shape of the correlation dimension.} 
 \label{fig:D2_models}
\end{figure}

To provide insight into the model dependence of the correlation dimension, we plot how it varies for each of these free parameters in Figure~\ref{fig:D2_models}. In each case, we vary a single parameter, otherwise keeping the model fixed at the `matter' values and at $z=0$ for our fiducial cosmology. The galaxy bias parameters have the largest impact on the homogeneity scale, effectively shifting the correlation dimension to larger scales as they are increased --- as galaxies become more biased they act as increasingly `special' tracers of the underlying density field and make the sample look less and less homogeneous. The second-order and stochastic bias parameters $b_{2}\sigma_{8}$ and $b_{s}\sigma_{8}$ have very similar effects and mostly enter at quadratic order in the model --- i.e., the curves for $b_{2}\sigma_{8} = -2$ and $b_{2}\sigma_{8} = 2$ are very similar on the plot, with difference between values of similar magnitude, and between $b_{2}\sigma_{8}$ and $b_{s}\sigma_{8}$ only apparent on small scales. 

The counterterm $\alpha_{0}$ and stochastic term $\mathrm{SN}_{0}$ behave somewhat differently --- as these mostly act on the non-linear power spectrum, they modulate the shape of the correlation dimension on small-scales and around the BAO scale, but have weaker impact on the homogeneity scale. Another notable feature of these parameters is their ability to cause an upturn in the correlation dimension at $r<50h^{-1}$Mpc. This is ultimately caused where galaxies are clustered in such a way that the correlation function is larger than it's volume-average (e.g., it is increasing with radius or only slowly decreasing), which occurs generically in the correlation function due to the presence of non-linearities. However, for particularly extreme choices of the counter and stochastic terms, this behavior is expressed on scales beyond those typically associated with non-linearities and within those plotted in Figure~\ref{fig:D2_models}.

In our fits, particularly to the QSO sample, we found that it was necessary to include all of the five terms as free parameters in order to model the correlation dimension across the full range of scales and to provide an accurate reconstruction of the matter correlation dimension.

\subsubsection{Mock validation}

\begin{figure*}[]
\centering
    \includegraphics[scale=0.53]{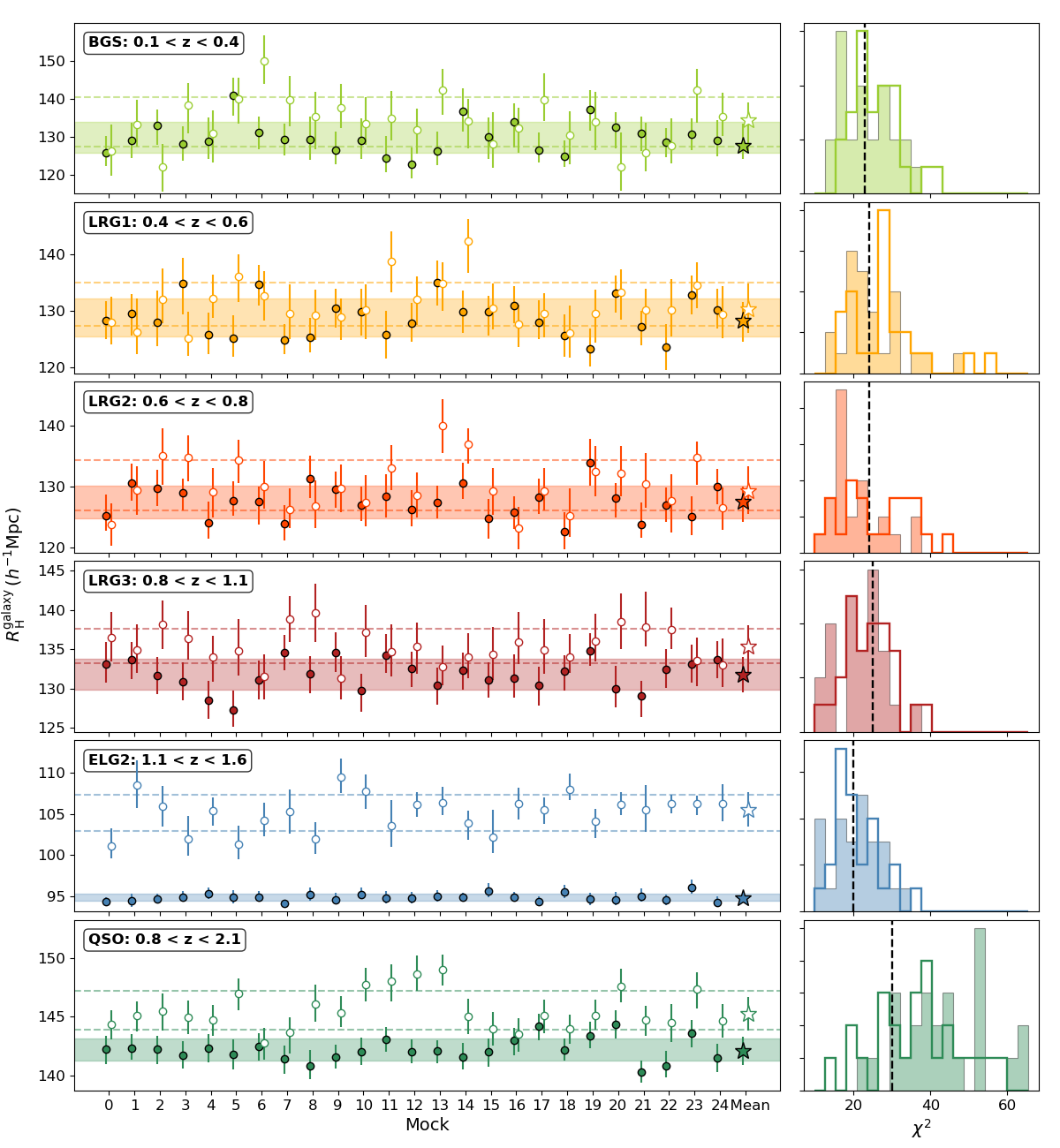}
	\caption{Validation of the model-fitting procedure via constraints on the homogeneity scale of galaxies measured from Abacus-2 simulations for each DESI tracer. Closed and open points show constraints from the NGC and SGC sky patches respectively, with mock number labeled on the x-axis and the last, `star', data point showing the fit to the mock mean using the covariance for a single realisation (i.e., the same covariance matrix as used for fits to the data). Horizontal bands show the standard deviation, centred on the mean, across the individual mock fits (filled band for NGC and dashed lines for SGC respectively). Note that the star, coming from fits to the mean data vector, and the center of the horizontal band, coming from the mean of the individual fits, need not agree --- the good agreement hence validates our approach. The right-hand panels for each tracer show the best-fit $\chi^{2}$ for each of the individual fits, with filled and unfilled histograms used for the NGC and SGC sky patches respectively. The vertical dashed line shows the number of degrees of freedom for each tracer. In all cases, we see consistent differences in the results from the NGC and SGC for a given tracer across individual mocks and the mock mean, as would be expected if this was due to the fiber assignment process. Nonetheless, the results for the mock mean are consistent with the scatter between mocks and the $\chi^{2}$ for all but the QSO results are close to the number of degrees of freedom, indicating our estimated covariance matrices (and associated rescaling factors in Table~\ref{tab:rescaling}) error bars are reliable. The high QSO $\chi^{2}$ is likely due to non-linear modelling inaccuracies, but this was not found in the fits to the real data.} 
 \label{fig:mock_desilike}
\end{figure*}

To validate our approach and the ability of this physical model (which is generally less flexible than a basis-spline) to fit the data to high accuracy, we first ran fits to each of the individual Abacus-2 simulations using the same EZmock covariance as for the data. We did this for all tracers, comparing the homogeneity scale between the different individual mocks with the mock mean. We also compared the recovered best-fit $\chi^{2}$ for each fit with the expectation given our numbers of degrees of freedom. The results are shown in Figure~\ref{fig:mock_desilike}.

The most apparent feature in this plot is the same difference between the NGC and SGC sky patches identified in our model-independent fits. Beyond this, we find good consistency between the fits to individual mocks and to the mock mean (using the covariance matrix for a single realisation rather than the error on the mean). For all tracers, the mean homogeneity scale recovered from each set of $25$ individual realisations is consistent with that obtained from the fit to the mean correlation dimension. The scatter between the maximum \textit{a posteriori} (MAP) values for the individual fits is also largely consistent with the average uncertainty from those same fits, indicating that our covariance matrix with rescaling factors from Table~\ref{tab:rescaling} is reliable. 

Figure~\ref{fig:mock_desilike} also shows the best-fit $\chi^{2}$ values for each of the individual mock fits. In general, we find that the $\chi^{2}$ values for most of the tracers are reasonable compared to the number of degrees of freedom. The exception is for the QSOs, where the average $\chi^{2}$ is relatively large --- we attribute this to the wide range of scales necessary to use in the fit to encapsulate both the galaxy and (as will be demonstrated later) matter homogeneity scales for this data, coupled with the relatively small statistical uncertainty. These combined present difficulties for the model in accurately capturing nonlinearities on small scales and around the BAO feature. However, in fitting the real data, we find that the $\chi^{2}$ is actually very reasonable. As such, we conclude that our fitting methodology is robust enough for application to the real data, and leave further investigation of how the modelling could be improved for the QSO sample to future work.

\subsubsection{Real data}

Results from our model-dependent fits to the real DESI data are summarised in Table~\ref{tab:model-dependent}. As for our model-independent fits, we report separate constraints for the NGC and SGC sky patches for each tracer and compare these to the constraints using the Abacus-2 mock means, which represent our $\Lambda$CDM baseline and which we expect to tend to homogeneity on large scales.

\begin{table*}
 \caption{Constraints on the \textit{galaxy} homogeneity scale from model-dependent fits. Columns are as described in Table~\ref{tab:model-independent}. Our fits to the data are generally robust (with the high LRG1 SGC $\chi^{2}$ likely due to the particular noise properties of the data) and consistent with the $\Lambda$CDM Abacus-2 mock means. The exceptions are small tensions in the QSO and LRG3 bins --- as argued in the main text, for the QSO results this may caused by modelling inaccuracies when fitting the mocks, but for the LRG3 results this may be evidence of a faster transition to homogeneity in the real data than expected, which is worth investigating further with more complete DESI data.}\label{tab:model-dependent}
  \centering
 {\tablefont\begin{tabular}{@{\extracolsep{\fill}}lcccccccccc}
   \toprule
    \multirow{3}{*}{Name} & \multirow{3}{*}{$z_{\mathrm{eff}}$} & \multirow{3}{*}{Fitting Range} & \multicolumn{2}{c}{Abacus-2} & \multicolumn{4}{c}{Data} & \multicolumn{2}{c}{\multirow{2}{*}{$|\Delta R_{H}^{\mathrm{galaxy}}|$}}\\
      & & & \multicolumn{2}{c}{$R_{H}^{\mathrm{galaxy}}\,(h^{-1}\mathrm{Mpc})$} & \multicolumn{2}{c}{$R_{H}^{\mathrm{galaxy}}\,(h^{-1}\mathrm{Mpc})$} & \multicolumn{2}{c}{$\chi^{2}$/dof} & &  \\
      & & & NGC & SGC & NGC & SGC & NGC & SGC & NGC & SGC \\
     \hline
    BGS & 0.295 & $48h^{-1}\mathrm{Mpc} < r < 160h^{-1}\mathrm{Mpc}$ & $131.0^{+1.2}_{-1.0}$ & $136.9^{+1.5}_{-1.8}$ & $131.6^{+4.7}_{-5.4}$ & $127.1^{+6.8}_{-4.6}$ & $22.3/23$ & $14.2/23$ & $0.11\sigma$ & $1.39\sigma$\\
    LRG1 & 0.510 & $44h^{-1}\mathrm{Mpc} < r < 160h^{-1}\mathrm{Mpc}$ & $130.2^{+0.9}_{-0.9}$ & $132.1^{+1.1}_{-1.2}$ & $128.9^{+3.4}_{-3.4}$ & $134.2^{+4.1}_{-4.6}$ & $14.2/24$ & $39.3/24$ & $0.37\sigma$ & $0.44\sigma$\\
    LRG2 & 0.706 & $44h^{-1}\mathrm{Mpc} < r < 160h^{-1}\mathrm{Mpc}$ & $128.7^{+0.8}_{-0.8}$ & $132.3^{+0.9}_{-1.0}$& $127.1^{+3.4}_{-3.2}$ & $129.7^{+4.4}_{-4.5}$ & $24.6/24$ & $23.1/24$ & $0.46\sigma$ & $0.58\sigma$ \\
    LRG3 & 0.919 & $40h^{-1}\mathrm{Mpc} < r < 160h^{-1}\mathrm{Mpc}$ & $132.8^{+0.6}_{-0.6}$ & $137.5^{+0.8}_{-0.7}$ & $128.7^{+2.3}_{-2.2}$ & $129.1^{+3.3}_{-2.8}$ & $25.3/25$ & $19.9/25$ & $1.72\sigma$ & $2.49\sigma$ \\
    ELG2 & 1.317 & $28h^{-1}\mathrm{Mpc} < r < 132h^{-1}\mathrm{Mpc}$ & $94.9^{+0.2}_{-0.2}$ & $105.3^{+0.5}_{-0.5}$ & $94.1^{+0.5}_{-0.5}$ & $105.4^{+2.4}_{-2.1}$ & $12.5/20$ & $23.0/20$ & $1.49\sigma$ & $0.05\sigma$ \\
    QSO &  1.491 & $28h^{-1}\mathrm{Mpc} < r < 172h^{-1}\mathrm{Mpc}$ & $142.7^{+0.2}_{-0.2}$ & $146.1^{+0.3}_{-0.3}$ & $141.0^{+1.3}_{-1.1}$ & $142.6^{+1.4}_{-1.4}$ & $35.8/30$ & $31.1/30$ & $1.29\sigma$ & $2.44\sigma$
   \botrule
    \end{tabular}}
\end{table*}

Looking at these results, we find a range of homogeneity scale measurements from the data that are qualitatively and quantitatively similar to our model-independent fits --- the ELG2 bin exhibits the lowest homogeneity scale, and the QSO the largest, with values $94.1^{+0.5}_{-0.5}h^{-1}$Mpc and $142.6^{+1.4}_{-1.4}h^{-1}$Mpc respectively. In general, the constraints from the model-dependent fits are tighter than those using basis-splines, as could be expected given the reduced modelling flexibility. Nonetheless, the $\chi^{2}$ values for all fits are quite reasonable, which the exception of the LRG1 SGC fits with $\chi^{2}=39.3$ for 24 degrees of freedom. This tracer also had a moderately high $\chi^{2}$ in our model-independent fits. However, in general, the model seems to perform well on average for the LRG1 mocks as shown in Figure~\ref{fig:mock_desilike}, while the data $\chi^{2}$ is consistent with the broader distribution of mock values (2 out of 25 of the Abacus-2 mocks exhibit higher $\chi^{2}$). As such, we attribute this to a combination of being `unlucky' with this particular dataset and small non-Gaussianities in the distribution of $D_{2}$ measurements for this tracer.

More interesting results appear in the comparison between the Abacus-2 mock mean and the data. When fitting using our cosmology-based model, we find stronger evidence of tension between the data and the mocks for the LRG3 and QSO tracers, with the SGC sky patch in particular showing discrepancies of $2.49\sigma$ and $2.44\sigma$ respectively. Comparing the results between Tables~\ref{tab:model-independent} and~\ref{tab:model-dependent}, we can see that the increased tension for the QSO results arises from upwards shifts in the mock mean measurements in the model-dependent compared to the model-independent fits. This shift is detected at $3.61\sigma$ and $1.97\sigma$ for the NGC and SGC respectively, given the small errors on the mock mean measurements. However, no such substantial shifts are seen in the QSO data (nor indeed in any of the other mock or data measurements). This is the cause of the increase in tension between the QSO mocks and the data. However, given that Figure~\ref{fig:mock_desilike} demonstrates some level of difficulty in getting the model to fit the mock measurements, we are unable to rule out that this shift is not a result of modelling inaccuracies.

For the LRG3 tension, again comparing Tables~\ref{tab:model-independent} and~\ref{tab:model-dependent}, this arises partially from a small downwards shift in the homogeneity scale from the model-dependent fits to the data compared to the model-independent ones (which is fully consistent with the errors bars and so not detected with any significance), but mostly from the large reduction in uncertainty when fitting with our cosmology-dependent model rather than more flexible basis-splines. There is no evidence that the fit to the data or mocks for the LRG3 bin is poor or unreliable. This tension is hence interesting and carries hints of a real discrepancy between the data and mocks that will be worth investigating with future DESI data. It is important to note, however, that the difference is such that the data exhibits a \textit{smaller} homogeneity scale and more \textit{rapid} transition to homogeneity than the simulations, and so this tension does not represent disagreement with the cosmological principle.

\subsubsection{Reconstructing the matter correlation dimension}
\label{sec:rescaling}
A benefit of the model-dependent fitting approach is that one can use the model with and without the effects of redshift-space distortions and galaxy bias to reconstruct the best-fitting \textit{matter} correlation dimension for the data, and the corresponding \textit{matter} homogeneity scale. This allows for more precise comparisons between different surveys and tracers. Having investigated the \textit{galaxy} homogeneity scale in the data, investigating the \textit{matter} homogeneity scale is our next port-of-call.

We do this following \cite{Ntelis2017}, defining
\begin{equation}
    \hat{N}^{\mathrm{matter}}(r) = \frac{N^{\mathrm{matter}}(r,\bm{b},f=0)-1}{N(r,\bm{b},f)-1} \times (\hat{N}(r) - 1) + 1, 
    \label{eq:matter_msc}
\end{equation}
where $\hat{N}(r)$ are the measured \textit{galaxy} mean scaled counts and $N(r,\bm{b},f)$ is the model which includes our set of bias parameters $\bm{b}$ and RSD controlled by our fiducial growth rate of structure $f$. $N^{\mathrm{matter}}(r,\bm{b},f=0)$ is the model for \textit{matter} obtained by setting $b_{1}\sigma_{8}=\sigma_{8}$ and $b_{2}\sigma^{2}_{8}=b_{s}\sigma^{2}_{8}=\alpha_{0}=\mathrm{SN}_{0}=0.0$, and where we also set the growth rate of structure $f=0$ to remove the effects of redshift-space distortions.\footnote{Note that some works such as \cite{Scrimgeour2012} and \cite{Laurent2016} assume a linear bias model, in which case the combination of Eqs.~\ref{eq:msc} and~\ref{eq:matter_msc} reduces to $\hat{N}^{\mathrm{matter}}(r) = (\hat{N}(r) - 1)/b_{1} + 1$ (where the $b_{1}$ parameter incorporates both the true galaxy bias and growth rate of structure which are degenerate in linear theory). \cite{Goncalves2021} instead apply a similar linear scaling directly to the correlation dimension $\hat{D}^{\mathrm{matter}}_{2}(r) = (\hat{D}_{2}(r) - 3)/b_{1} + 3$, which is further derived from Eq.~\ref{eq:D2} in the limit $|\hat{D}_{2}(r) - 3| << 1$. The approach we use has the benefit of avoiding these assumptions at the cost of a more complex and potentially model-dependent implementation.}

This approach effectively uses the scale-dependent ratio of the volume-averaged correlation function to rescale the data. The matter correlation dimension is then constructed from the rescaled $\hat{N}^{\mathrm{matter}}(r)$ in the same way as for the original mean scaled counts, and we fit the matter correlation dimension using another basis-spline to determine the crossing-point. Our constraints on the matter homogeneity scale are obtained by applying the above rescaling to both the data and all the EZmocks for 1000 randomly selected sets of model parameters drawn from the model-dependent MCMC chains after removing the burn-in. This results in 1000 rescaled data vectors and 1000 rescaled mock-based covariance matrices, each of which is then fit using the basis-spline model to extract 1000 samples of the matter homogeneity scale. The distribution of these matter homogeneity scale values is used to obtain our constraints. We found that our results were generally reliable so long as at least 200 MCMC parameter sets were used for the rescaling.

\begin{figure*}[]
\centering
    \includegraphics[scale=1.0]{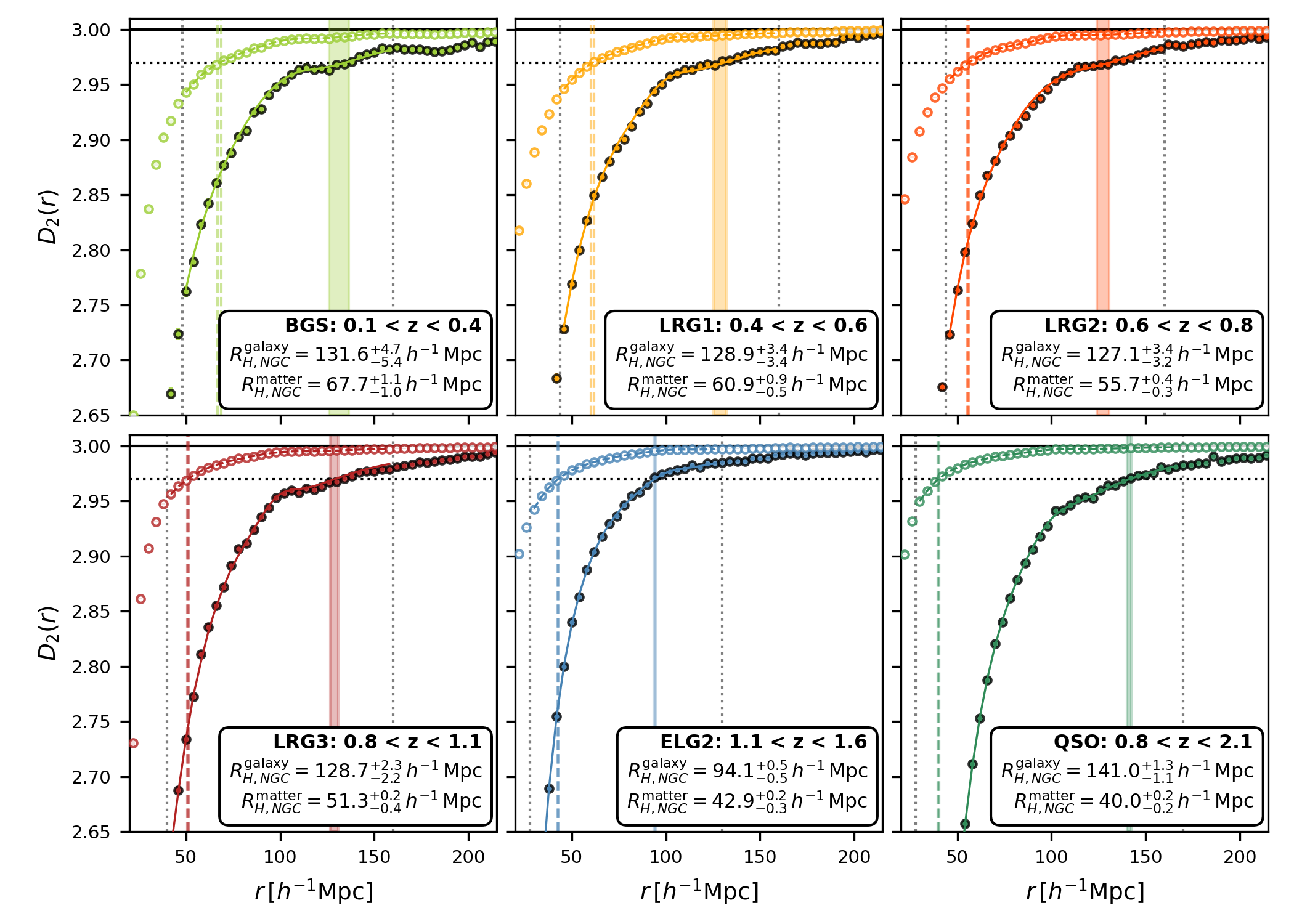}
	\caption{Measurements of the correlation dimension as a function of scale for each of the 6 DESI redshift bins considered in this work for the NGC. Closed points and solid lines show the measurements in the \textit{galaxies} and best-fitting, cosmology-dependent, \texttt{LPTVelocileptors} model. Open points and the dashed-colored line show the corresponding reconstruction of the \textit{matter} correlation dimension (using Equation~\ref{eq:matter_msc}) and best-fitting basis-spline model used to infer the matter homogeneity scale. The subsequent constraints on the homogeneity scales for the galaxies and matter in each panel are highlighted as the colored vertical band and pair of dashed lines for the galaxies and matter respectively, and also included as text. In all cases, the dotted vertical lines show the range of scales included in our fits, while the dotted and solid horizontal lines denote the 1\% homogeneity threshold $D_{2}(r)=2.97$, and the asymptotic homogeneity value $D_{2}(r)=3.0$ respectively. We see that the \textit{galaxy} homogeneity scale varies in a complicated way between tracers based on the effective redshift and galaxy bias of the sample, but that the matter homogeneity scale exhibits a monotonic decrease from the lowest-redshift BGS sample to the highest-redshift QSO sample.} 
 \label{fig:matter_homogeneity_NGC}
\end{figure*}

Our results from this procedure applied to the NGC data are demonstrated in Figure~\ref{fig:matter_homogeneity_NGC} and summarised in Table~\ref{tab:matter_homogeneity} for both galactic caps. The plot for the SGC is qualitatively similar to that for the NGC and so not shown here.

Our reconstruction finds a matter homogeneity scale that is well-fit by the basis-spline model and decreases monotonically with increasing redshift --- structures grow as redshift decreases, leading to more inhomogeneities and a larger scale of homogeneity. Despite the differences in the homogeneity scales between the NGC and SGC sky patches identified earlier, the \textit{matter} homogeneity scale shows much better agreement. Differences we previously attributed to fiber assignment are now absorbed into differences in the free galaxy bias, shot-noise, and counter-term parameters, leading to greater consistency in the homogeneity scale.  

We also find good consistency between the $\Lambda$CDM expectations from our fiducial cosmology and the reconstructed homogeneity scale from the Abacus-2 mock means, however there is evidence of small modelling systematics in the LRG1, ELG2, and QSO bins, which are detected at particularly high significance for these last two cases (e.g., at $>10\sigma$ in the case of the QSOs) due to the high precision of the mock mean measurements. These systematic differences (i.e., the numerical differences between the second and third, or second and fourth columns, of Table~\ref{tab:matter_homogeneity}) are treated as systematic errors as a result of non-linear modelling inaccuracies, and included in quadrature alongside the statistical errors on the data. As such the values for the data reported in Table~\ref{tab:matter_homogeneity} and used hereafter include both statistical and systematic error contributions. Despite this difference being detected at high significance, it is important to note that the systematic errors applied to the \textit{data} for all tracers are subdominant, or at most comparable, to the statistical errors.

\begin{figure}[]
\centering
    \includegraphics[scale=0.75]{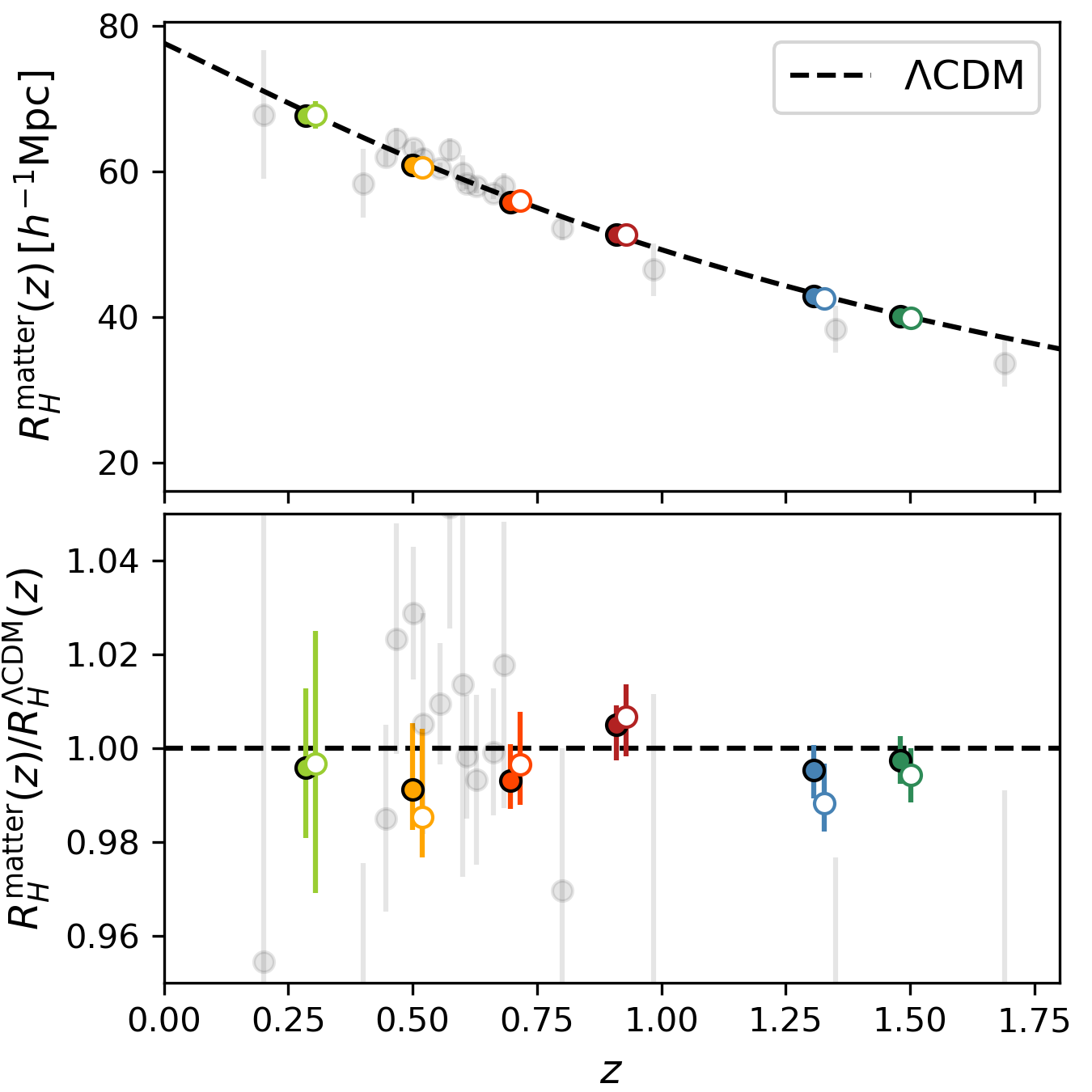}
	\caption{Constraints on the matter homogeneity scale as a function of redshift. Coloured points show our new DESI DR1 constraints, following the same colour scheme as used elsewhere in this work and placed at the effective redshift of the tracer. Closed and open points show NGC and SGC sky patches respectively. Faint grey points are results from previous works using WiggleZ \citep{Scrimgeour2012} and SDSS data \citep{Ntelis2017, Goncalves2018}, rescaled to account for differences in the fiducial cosmology used in those works and here. The dashed line is the predicted homogeneity scale for our fiducial $\Lambda$CDM model, while the bottom panel plots ratios relative to this model.} 
 \label{fig:rh_vs_redshift}
\end{figure}

Figure~\ref{fig:rh_vs_redshift} now plots the constraints on the matter homogeneity scale as a function of redshift against our fiducial model prediction and other literature constraints \citep{Scrimgeour2012, Ntelis2017, Goncalves2018}.\footnote{As discussed in \cite{Ntelis2018}, constraints on the homogeneity scale depend on the assumed cosmology used to convert galaxy redshifts to cartesian coordinates. As such, for the purposes of this comparison, we have rescaled the results from previous works to our fiducial cosmology.} Our measured homogeneity scales are in broad agreement with those from previous datasets, albeit with MAP values slightly larger at $z>0.75$ and closer to the $\Lambda$CDM prediction, and with the DESI data offering substantially smaller uncertainties. Our results at all redshifts closely track the $\Lambda$CDM prediction, although tend to lie slightly below it. It is interesting to note that the volume averaged BAO distance constraints from \cite{DESI2024III}, using the same data and fiducial cosmology, also lie slightly below the $\Lambda$CDM expectations. This may point towards a similar slight preference for a different matter density or dark energy equation of state than our fiducial cosmology, although we leave further investigation of this to future work.     

\begin{table}
 \caption{Constraints on the \textit{matter} homogeneity scale from model-dependent fits and the rescaling procedure described in Section~\ref{sec:rescaling}. The first three columns show the name and effective redshift for each tracer sample, and the homogeneity scale expected for our fiducial cosmology. We then show homogeneity scale constraints for the NGC and SGC sky patches for the Abacus-2 mock mean (with covariance matrix divided by 25 to reflect the error on the mean) and real data. Fitting ranges for these constraints match those in Table~\ref{tab:model-dependent}. Uncertainties reported for the data include both statistical and systematic contributions, with the latter obtained comparing the columns for Abacus-2 against the $\Lambda$CDM predictions.}\label{tab:matter_homogeneity}
  \centering
 {\tablefont\begin{tabular}{@{\extracolsep{\fill}}lcccccc}
   \toprule
    \multirow{3}{*}{Name} & \multirow{3}{*}{$z_{\mathrm{eff}}$} & \multirow{3}{0.8cm}{\centering \shortstack{$R_{H}^{\Lambda\mathrm{CDM}}$\\$(h^{-1}\mathrm{Mpc})$}} & \multicolumn{2}{c}{Abacus-2} & \multicolumn{2}{c}{Data} \\
      & & & \multicolumn{2}{c}{$R_{H}^{\mathrm{matter}}\,(h^{-1}\mathrm{Mpc})$} & \multicolumn{2}{c}{$R_{H}^{\mathrm{matter}}\,(h^{-1}\mathrm{Mpc})$} \\
      & & & NGC & SGC & NGC & SGC \\
     \hline
    BGS & $0.295$ & $67.96$ & $67.94^{+0.28}_{-0.15}$ & $67.74^{+0.42}_{-0.24}$ & $67.7^{+1.1}_{-1.0}$ & $67.7^{+1.9}_{-1.9}$ \\
    LRG1 & $0.510$ & $61.43$ & $61.22^{+0.18}_{-0.08}$ & $61.33^{+0.14}_{-0.23}$ & $60.9^{+0.9}_{-0.5}$ & $60.5^{+1.2}_{-0.5}$ \\
    LRG2 & $0.706$ & $56.13$ & $56.12^{+0.09}_{-0.09}$ & $56.05^{+0.14}_{-0.09}$& $55.7^{+0.4}_{-0.3}$ & $55.9^{+0.6}_{-0.5}$ \\
    LRG3 & $0.919$ & $51.01$ & $51.01^{+0.07}_{-0.04}$ & $51.12^{+0.07}_{-0.08}$ & $51.3^{+0.2}_{-0.4}$ & $51.4^{+0.4}_{-0.4}$ \\
    ELG2 & $1.317$ & $43.05$ & $42.86^{+0.03}_{-0.04}$ & $42.83^{+0.05}_{-0.04}$ & $42.9^{+0.2}_{-0.3}$ & $42.5^{+0.4}_{-0.3}$ \\
    QSO & $1.491$ & $40.14$ & $39.95^{+0.01}_{-0.02}$ & $39.93^{+0.02}_{-0.03}$ & $40.0^{+0.2}_{-0.2}$ & $39.9^{+0.2}_{-0.2}$ 
   \botrule
    \end{tabular}}
\end{table}

\section{CONCLUSIONS}
\label{sec:conclusion}
In this work, we have presented the most precise constraints to date of the scales at which the Universe reaches within $1\%$ of homogeneity between $0.1 < z < 2.1$. Using the first data release (DR1) from the Dark Energy Spectroscopic Instrument (DESI), we have measured the mean scaled counts-in-spheres and the correlation dimension in 6 different redshift bins and two different sky patches, north and south of the Galactic plane. In all cases, we have found that the galaxy samples clearly asymptote to the expectations for a homogeneous Universe, reaching within $1\%$ homogeneity on scales between $90-145h^{-1}$Mpc depending on the exact redshift and galaxy bias of the tracer.

As part of our analysis, we identified a small but significant difference (up to $\sim10\%$) in the homogeneity scale measured between DESI's North and South Galactic Cap data sets. We found this difference was also present in our $\Lambda$CDM simulations of the DESI tracers, which include the same selection and observational effects. Further investigation using ``complete'' mocks demonstrated this is due to small-scale fiber assignment/collision issues and the current large completeness variations across the DESI DR1 footprint (as can be seen in Figure~\ref{fig:desi_sky}). We expect this to be improved in future datasets as the completeness improves. It could also potentially be ameliorated using a method akin to the $\theta$-cut process \citep{Pinon2025} used to correct the DESI small-scale clustering in \cite{DESI2024III} and \cite{DESI2024V}, or more sophisticated `pairwise inverse-probability' weights than we used here \citep{Bianchi2025}.

In our work, we accounted for the NGC/SGC differences by performing comparisons between the data and simulations for each sky patch separately. Doing so, we find no evidence for any deviations from the cosmological principle in our dataset. The only significant differences between the homogeneity scale measured from the data and $\Lambda$CDM simulations occur for the QSO and LRG3 tracers (at $2.44$ and $2.49\sigma$ respectively) when fitting with a cosmology-dependent model --- results which may also go away with improved non-linear modelling and more complete DESI data; and which in any case denote a \textit{more} rapid transition to homogeneity in the data than predicted by the simulations. This may have implications for understanding cosmic dipole and bulk flow tensions (e.g., \citealt{Whitford2023, Oayda2024, LandStrykowski2025})  --- the DESI data does not find evidence of large departures from homogeneity or isotropy that, at a glance, would appear to be needed to confirm these results. However, further work is required to understand how the differences in approaches between the measurement techniques used in this work and in past dipole/bulk flow analyses may impact their comparability. 

Using our cosmology-dependent model, we have reconstructed the correlation dimension of the underlying \textit{matter} field at each of our 6 redshifts, demonstrating a homogeneity scale in the matter field that decreases with increasing redshift, in excellent agreement with the predictions of our fiducial $\Lambda$CDM model, albeit with a slight preference for smaller scales. This could indicate a small difference between our fiducial cosmological model and the model preferred by the data (which is also found in the DESI BAO and RSD constraints using the same data; \citealt{DESI2024VII, DESI2024VI}), although we leave exploration of our results and their potential use as a standard ruler to future studies.

Lastly, we end with a discussion of the recent results from \cite{SylosLabini2025}, which were published while this manuscript was being prepared but also use the DESI DR1 data. They attempt to use volume-limited samples to identify whether or not the data exhibits a transition to homogeneity, finding, in contrast to what we presented here, that \textit{no} such transition is reached within $400h^{-1}$Mpc for neither the BGS nor LRG samples. The reason for the difference boils down to the questions of 1) whether our use of a random catalogue to normalise the mean scaled counts guarantees us to find homogeneity even for inhomogeneous distributions and 2) whether or not the DESI DR1 data can be used to construct robust volume-limited samples.

In order to address these questions we have presented exercises using several different toy models, including homogeneous and inhomogeneous distributions. We have found that, contrary to the claims of \cite{SylosLabini2025}, the use of a random catalogue does not guarantee one to find homogeneity. Even in the case where the redshift distribution of the randoms is sampled directly from the data (as is done for the DESI DR1 large-scale structure catalogues, and has been done in the past for analyses of WiggleZ and Sloan Digital Sky Survey data \citealt{Scrimgeour2012,Ntelis2017,Goncalves2018}), we are still able to identify inhomogeneous distributions such as fractal, Gaussian, and Giant Void universes. The exception to this is when the observer happens to sit in a particularly lucky (or unlucky, depending on your mindset) location in an inhomogeneous distribution such that they still measure an approximately isotropic distribution of sources (e.g., at the centre of the aforementioned Gaussian or Giant Void distributions).

On the flip side, we have demonstrated that the current DESI DR1 data is subject to significant completeness variations: in the angular and radial directions, on small and large scales, and due to the complex combination of how DESI targets are selected, assigned fibers, and observed. The impact of these is such that even for our $\Lambda$CDM simulations, which must approach homogeneity and isotropy on large-scales, and when using our random catalogues and systematic weights that should partially mitigate completeness variations, we found significant differences between the homogeneity scales measured in the Northern and Southern Galactic Caps. This highlights the difficulty in defining robust volume-limited samples from the current DR1 data and the risk of over-estimating the level of inhomogeneity in the data due to unmodelled selection effects. Although we don't reproduce the methods of \cite{SylosLabini2025} (or recent followup work in \citealt{SylosLabini2026}), we suspect that these unmodelled selection effects are responsible for the apparent lack of large-scale homogeneity in their results, while our use of a random catalogue corrects for this.

For these two reasons we believe that using a random catalogue is the best way to analyse the current DESI data, and that doing so gives robust evidence in support of the cosmological principle. Although we argue a volume-limited sample cannot be constructed accurately with DR1, future DESI data with more passes may allow for the construction of truly volume-limited samples with which to test consistency against results using the random catalogue approach. Future data also offers the promise of tighter and more accurate constraints on the homogeneity scale across a wider range of redshifts, and with finer tomographic binning.  

\section*{Data Availability}
Data Release 1 of the Dark Energy Spectroscopic Instrument used for this work can be found at \url{https://data.desi.lbl.gov/doc/releases/dr1/}. Codes necessary to reproduce the analysis in this work are available on Github: \url{https://github.com/cullanhowlett/homogeneity}, while all derived data products including measurements of the mean scaled counts and MCMC chains can be found here: \url{https://doi.org/10.5281/zenodo.23027000}

\section*{Acknowledgements}
CH, LG, and TD acknowledge support from the Australian Government through the Australian Research
Council’s Discovery Project (projects DP220101395 and DP250102917) and Centre of Excellence (CE) funding schemes.

This work acknowledges the use of Anthropic’s Claude generative AI models (Claude Sonnet 4.6 and later versions) in developing and refactoring analysis and plotting code, preparing code documentation, and in preparing this manuscript. All AI-assisted outputs were reviewed, tested and verified by the authors, who take full responsibility for the correctness and originality of the work.

This material is based upon work supported by the U.S. Department of Energy (DOE), Office of Science, Office of High-Energy Physics, under Contract No. DE–AC02–05CH11231, and by the National Energy Research Scientific Computing Center, a DOE Office of Science User Facility under the same contract. Additional support for DESI was provided by the U.S. National Science Foundation (NSF), Division of Astronomical Sciences under Contract No. AST-0950945 to the NSF’s National Optical-Infrared Astronomy Research Laboratory; the Science and Technology Facilities Council of the United Kingdom; the Gordon and Betty Moore Foundation; the Heising-Simons Foundation; the French Alternative Energies and Atomic Energy Commission (CEA); the National Council of Humanities, Science and Technology of Mexico (CONAHCYT); the Ministry of Science, Innovation and Universities of Spain (MICIU/AEI/10.13039/501100011033), and by the DESI Member Institutions: \url{https://www.desi.lbl.gov/collaborating-institutions}. Any opinions, findings, and conclusions or recommendations expressed in this material are those of the author(s) and do not necessarily reflect the views of the U. S. National Science Foundation, the U. S. Department of Energy, or any of the listed funding agencies.

The authors are honored to be permitted to conduct scientific research on I'oligam Du'ag (Kitt Peak), a mountain with particular significance to the Tohono O’odham Nation.

\bibliographystyle{aa}
\bibliography{massive.bib}

\appendix
\label{app:binning}

\section{Robustness to fitting choices}

\setcounter{figure}{9}

\begin{figure}[]
\centering
    \includegraphics[scale=0.6]{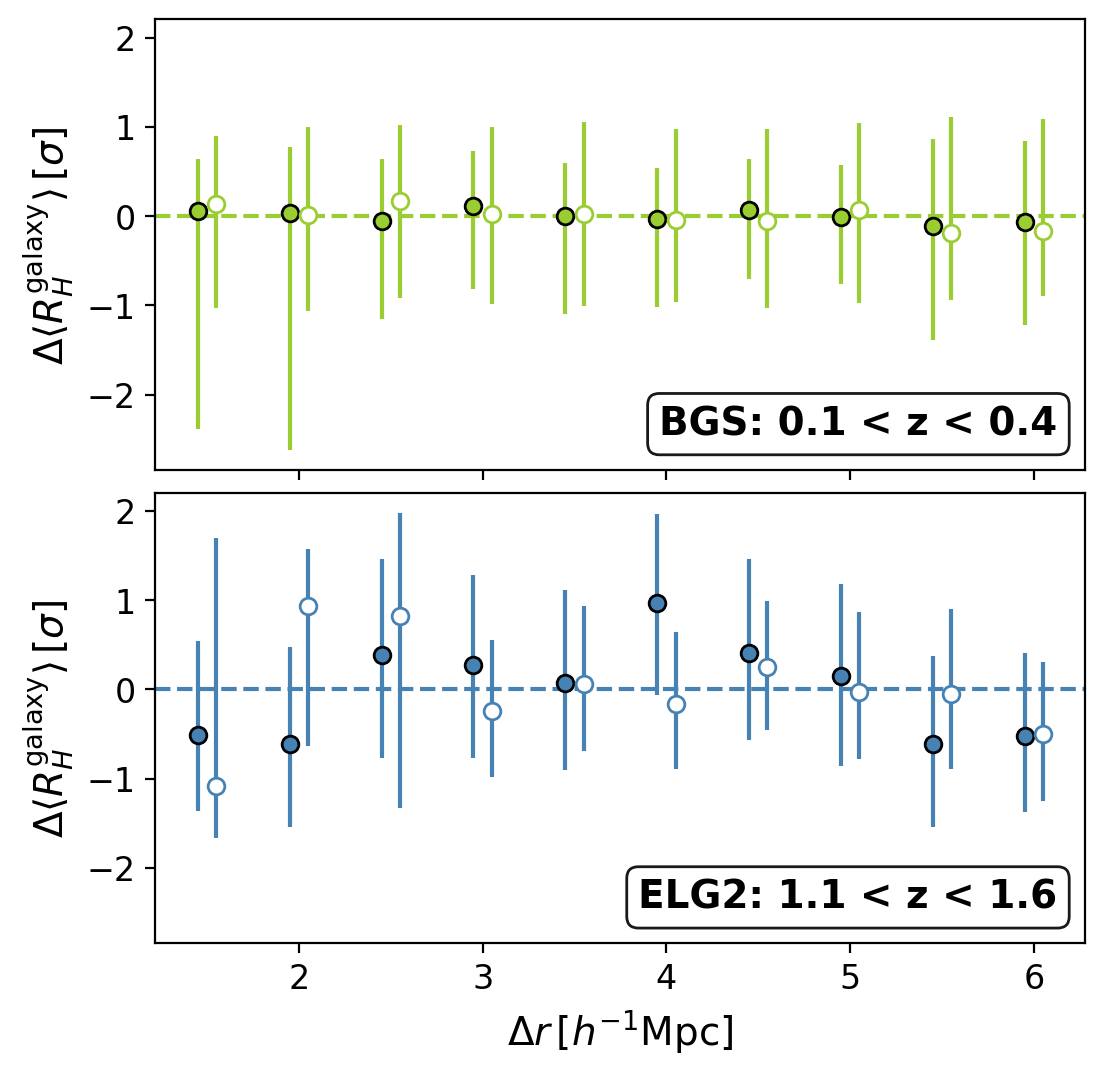}
	\caption{Variations in the measured homogeneity scale from spline-based fits as a function of measurement bin width. The upper and lower panels show results for the BGS and ELG2 tracer bins respectively, while closed points represent fits to NGC data, and open points fits to the SGC data. Each panel shows the difference in the homogeneity scale measured from the data compared to the average across all binning choices, and normalised by the average uncertainty on the measurements. The results in the main body of this paper were computed using $\Delta r = 4h^{-1}$Mpc, but variations as a result of different binning choices are within $1\sigma$.} 
 \label{fig:binning_tests}
\end{figure}

For all the analysis in this work, we have used a fixed bin size of $4h^{-1}$Mpc with which to measure the mean scaled counts (i.e., we measure the counts-in-spheres using a sphere radius that increases in steps of $4h^{-1}$Mpc) and take its derivative to obtain the correlation dimension. We settled on this choice as it offers a reasonable number of data points and size of covariance matrix compared to the number of degrees of freedom in our models and number of EZmocks. It also matches the correlation function binning used in other DESI works \citep{DESI2024II, DESI2024III, DESI2024V}. 

In Figure~\ref{fig:binning_tests} we demonstrate that our spline fits to the BGS and ELG2 data, which exhibit some of the largest and smallest uncertainties on the homogeneity scale respectively, change by at most $1\sigma$ for alternative choices of bin width between $1.5 - 6h^{-1}$Mpc.

\begin{figure*}[]
\centering
    \includegraphics[scale=0.7]{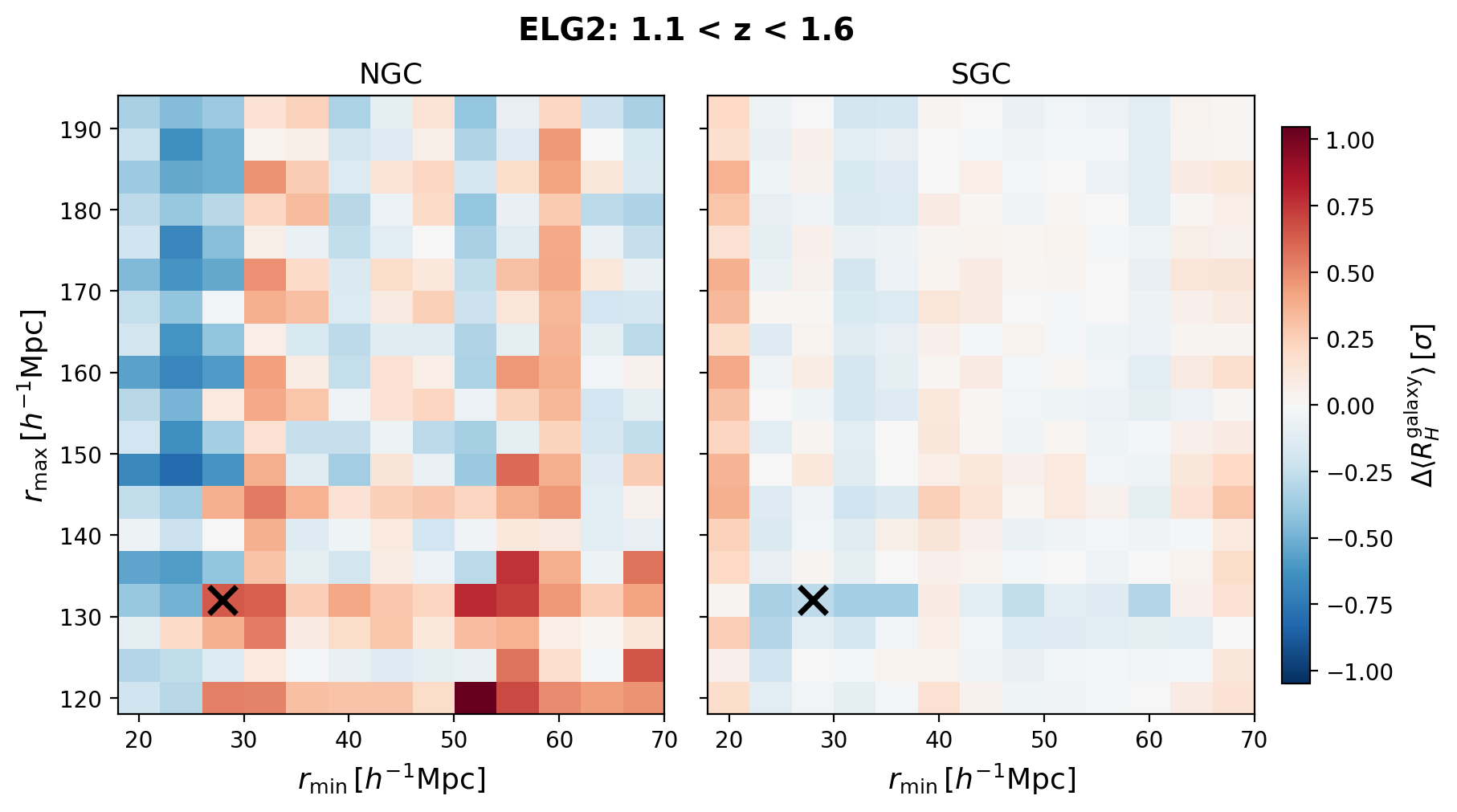}
	\caption{Variations in the measured homogeneity scale from spline-based fits as a function of the minimum and maximum scales included in the fitting for the ELG2 tracer bin. The left and right panels are for the NGC and SGC sky caps respectively. Each grid point is coloured based on the difference in the homogeneity scale measured from the data compared to the average across all choices of scale-cut, normalised by the average uncertainty across all the measurements. The cross marks the fiducial scale cuts used in this work, but the results remain within $\sim1\sigma$ across all the scale cuts tested, while other tracers show even weaker variations.} 
 \label{fig:scale_tests}
\end{figure*}

Furthermore, Figure~\ref{fig:scale_tests} demonstrates that the ELG2 measurements are robust to wide variations in the minimum and maximum scales used in the fitting, again with maximum variations in the homogeneity scale of $\sim1\sigma$. Similar tests for the other tracers show even smaller variation with the choice of scale cuts.

\section*{Affiliations}
$^{1}$ School of Mathematics and Physics, University of Queensland, Brisbane, QLD 4072, Australia.\\
$^{2}$ Centre for Astrophysics \& Supercomputing, Swinburne University of Technology, P.O. Box 218, Hawthorn, VIC 3122, Australia.\\
$^{3}$ Lawrence Berkeley National Laboratory, 1 Cyclotron Road, Berkeley, CA 94720, USA.\\
$^{4}$ Department of Physics, Boston University, 590 Commonwealth Avenue, Boston, MA 02215 USA.\\
$^{5}$ Leinweber Center for Theoretical Physics, University of Michigan, 450 Church Street, Ann Arbor, Michigan 48109-1040, USA.\\
$^{6}$ University of Michigan, 500 S. State Street, Ann Arbor, MI 48109, USA.\\
$^{7}$ IRFU, CEA, Universit\'{e} Paris-Saclay, F-91191 Gif-sur-Yvette, France.\\
$^{8}$ Department of Physics \& Astronomy, University of Rochester, 206 Bausch and Lomb Hall, P.O. Box 270171, Rochester, NY 14627-0171, USA.\\
$^{9}$ Institute for Astronomy, University of Edinburgh, Royal Observatory, Blackford Hill, Edinburgh EH9 3HJ, UK.\\
$^{10}$ Dipartimento di Fisica ``Aldo Pontremoli'', Universit\`a degli Studi di Milano, Via Celoria 16, I-20133 Milano, Italy.\\
$^{11}$ INAF-Osservatorio Astronomico di Brera, Via Brera 28, 20122 Milano, Italy.\\
$^{12}$ Department of Physics \& Astronomy, University College London, Gower Street, London, WC1E 6BT, UK.\\
$^{13}$ Institute of Cosmology and Gravitation, University of Portsmouth, Dennis Sciama Building, Portsmouth, PO1 3FX, UK.\\
$^{14}$ Departamento de Astrof\'{\i}sica, Universidad de La Laguna (ULL), E-38206, La Laguna, Tenerife, Spain.\\
$^{15}$ Instituto de Astrof\'{\i}sica de Canarias, C/ V\'{\i}a L\'{a}ctea, s/n, E-38205 La Laguna, Tenerife, Spain.\\
$^{16}$ Institut d'Estudis Espacials de Catalunya (IEEC), c/ Esteve Terradas 1, Edifici RDIT, Campus PMT-UPC, 08860 Castelldefels, Spain.\\
$^{17}$ Institute of Space Sciences, ICE-CSIC, Campus UAB, Carrer de Can Magrans s/n, 08913 Bellaterra, Barcelona, Spain.\\
$^{18}$ Instituto de F\'{\i}sica, Universidad Nacional Aut\'{o}noma de M\'{e}xico,  Circuito de la Investigaci\'{o}n Cient\'{\i}fica, Ciudad Universitaria, Cd. de M\'{e}xico  C.~P.~04510,  M\'{e}xico.\\
$^{19}$ NSF NOIRLab, 950 N. Cherry Ave., Tucson, AZ 85719, USA.\\
$^{20}$ Department of Astronomy \& Astrophysics, University of Toronto, Toronto, ON M5S 3H4, Canada.\\
$^{21}$ Department of Physics \& Astronomy and Pittsburgh Particle Physics, Astrophysics, and Cosmology Center (PITT PACC), University of Pittsburgh, 3941 O'Hara Street, Pittsburgh, PA 15260, USA.\\
$^{22}$ Institut de Ci\`encies del Cosmos (ICCUB), Universitat de Barcelona (UB), c. Mart\'i i Franqu\`es, 1, 08028 Barcelona, Spain.\\
$^{23}$ Departamento de F\'isica, Universidad de los Andes, Cra. 1 No. 18A-10, Edificio Ip, CP 111711, Bogot\'a, Colombia.\\
$^{24}$ Observatorio Astron\'omico, Universidad de los Andes, Cra. 1 No. 18A-10, Edificio H, CP 111711 Bogot\'a, Colombia.\\
$^{25}$ University of Virginia, Department of Astronomy, Charlottesville, VA 22904, USA.\\
$^{26}$ Fermi National Accelerator Laboratory, PO Box 500, Batavia, IL 60510, USA.\\
$^{27}$ Department of Astronomy, University of Texas at Austin, 2515 Speedway, TX 78712, USA.\\
$^{28}$ Center for Cosmology and AstroParticle Physics, The Ohio State University, 191 West Woodruff Avenue, Columbus, OH 43210, USA.\\
$^{29}$ Department of Physics, The Ohio State University, 191 West Woodruff Avenue, Columbus, OH 43210, USA.\\
$^{30}$ The Ohio State University, Columbus, 43210 OH, USA.\\
$^{31}$ Department of Physics, University of Michigan, 450 Church Street, Ann Arbor, MI 48109, USA.\\
$^{32}$ Department of Physics, The University of Texas at Dallas, 800 W. Campbell Rd., Richardson, TX 75080, USA.\\
$^{33}$ Department of Physics and Astronomy, University of California, Irvine, 92697, USA.\\
$^{34}$ Sorbonne Universit\'{e}, CNRS/IN2P3, Laboratoire de Physique Nucl\'{e}aire et de Hautes Energies (LPNHE), FR-75005 Paris, France.\\
$^{35}$ Departament de F\'{i}sica, Serra H\'{u}nter, Universitat Aut\`{o}noma de Barcelona, 08193 Bellaterra (Barcelona), Spain.\\
$^{36}$ Institut de F\'{i}sica d’Altes Energies (IFAE), The Barcelona Institute of Science and Technology, Edifici Cn, Campus UAB, 08193, Bellaterra (Barcelona), Spain.\\
$^{37}$ Instituci\'{o} Catalana de Recerca i Estudis Avan\c{c}ats, Passeig de Llu\'{\i}s Companys, 23, 08010 Barcelona, Spain.\\
$^{38}$ Department of Physics and Astronomy, Siena University, 515 Loudon Road, Loudonville, NY 12211, USA.\\
$^{39}$ Instituto de Estudios Astrof\'isicos, Facultad de Ingenier\'ia y Ciencias, Universidad Diego Portales, Av. Ej\'ercito Libertador 441, Santiago, Chile.\\
$^{40}$ Department of Physics and Astronomy, University of Waterloo, 200 University Ave W, Waterloo, ON N2L 3G1, Canada.\\
$^{41}$ Perimeter Institute for Theoretical Physics, 31 Caroline St. North, Waterloo, ON N2L 2Y5, Canada.\\
$^{42}$ Waterloo Centre for Astrophysics, University of Waterloo, 200 University Ave W, Waterloo, ON N2L 3G1, Canada.\\
$^{43}$ Space Sciences Laboratory, University of California, Berkeley, 7 Gauss Way, Berkeley, CA  94720, USA.\\
$^{44}$ University of California, Berkeley, 110 Sproul Hall \#5800 Berkeley, CA 94720, USA.\\
$^{45}$ Instituto de Astrof\'{i}sica de Andaluc\'{i}a (CSIC), Glorieta de la Astronom\'{i}a, s/n, E-18008 Granada, Spain.\\
$^{46}$ Departament de F\'isica, EEBE, Universitat Polit\`ecnica de Catalunya, c/Eduard Maristany 10, 08930 Barcelona, Spain.\\
$^{47}$ Universit\'{e} Clermont-Auvergne, CNRS, LPCA, 63000 Clermont-Ferrand, France.\\
$^{48}$ Department of Physics and Astronomy, Sejong University, 209 Neungdong-ro, Gwangjin-gu, Seoul 05006, Republic of Korea.\\
$^{49}$ Queensland University of Technology,  School of Chemistry \& Physics, George St, Brisbane 4001, Australia.\\
$^{50}$ Abastumani Astrophysical Observatory, Tbilisi, GE-0179, Georgia.\\
$^{51}$ Department of Physics, Kansas State University, 116 Cardwell Hall, Manhattan, KS 66506, USA.\\
$^{52}$ CIEMAT, Avenida Complutense 40, E-28040 Madrid, Spain.\\
$^{53}$ Max Planck Institute for Extraterrestrial Physics, Gie\ss enbachstra\ss e 1, 85748 Garching, Germany.\\
$^{54}$ Space Telescope Science Institute, 3700 San Martin Drive, Baltimore, MD 21218, USA.\\
$^{55}$ National Astronomical Observatories, Chinese Academy of Sciences, A20 Datun Road, Chaoyang District, Beijing, 100101, P.~R.~China.\\

\end{document}